\documentclass{aa}  
\usepackage{amsmath}
\usepackage{graphicx}
\usepackage{color}
\usepackage{txfonts}

\newcommand{\cmsyr}{{\rm cm\,s^{-1}\,yr^{-1}}}
\newcommand{\kms}{{\rm km\,s^{-1}}}
\newcommand{\ms}{{\rm m\,s^{-1}}}
\newcommand{\cms}{{\rm cm\,s^{-1}}}
\usepackage{siunitx}
\usepackage{booktabs}
\usepackage{longtable}

\begin{document} 

   \title{The ESPRESSO Redshift Drift Experiment\thanks{Based on Guaranteed Time Observations collected at the European Southern Observatory under ESO programmes 110.247Q and 112.25K7 by the ESPRESSO Consortium.}}

   \subtitle{I. High-resolution spectra of the Lyman-$\alpha$ forest of QSO J052915.80-435152.0}

   \author{A. Trost \inst{1,2,3}\thanks{andrea.trost@inaf.it}, 
          C. M. J. Marques \inst{4,5,6}, 
          S. Cristiani\inst{1,3,7}, 
          G. Cupani\inst{1,7}, 
          S. Di Stefano\inst{1,2},
          V. D'Odorico\inst{1,7}, 
          F. Guarneri\inst{1, 8}, 
          C. J. A. P. Martins\inst{4,5}, 
          D. Milakovi\'c\inst{1,7}, 
          L. Pasquini\inst{9,10},
          R. G\'enova Santos\inst{11,12}, 
          P. Molaro\inst{1,7}, 
          M. T. Murphy\inst{13},
          N. J. Nunes\inst{14,15}, 
          T. M. Schmidt\inst{16},
          Y. Alibert\inst{17}, 
          K. Boutsia\inst{18}, 
          G. Calderone\inst{1},
          J. I. Gonz\'alez Hern\'andez\inst{11, 12}, 
          A. Grazian\inst{19}, 
          G. Lo Curto\inst{9}, 
          E. Palle\inst{11,12}, 
          F. Pepe\inst{16}, 
          M. Porru\inst{1},
          N. C. Santos\inst{5,6}, 
          A. Sozzetti\inst{20}, 
          A. Su\'arez Mascare\~no\inst{10,11}, 
          \and
          M. R. Zapatero Osorio\inst{21}
          }
    \institute{
    INAF -- Osservatorio Astronomico di Trieste, Via G.B. Tiepolo, 11, I-34143 Trieste, Italy
    \and 
    Dipartimento di Fisica dell’Università di Trieste, Sezione di Astronomia, Via G.B. Tiepolo, 11, I-34143 Trieste, Italy
    \and
    INFN -- National Institute for Nuclear Physics, via Valerio 2, I-34127 Trieste
    \and
    Centro de Astrofísica da Universidade do Porto, Rua das Estrelas, 4150-762 Porto, Portugal
    \and
    Instituto de Astrofísica e Ciências do Espaço, Universidade do Porto, Rua das Estrelas, 4150-762 Porto, Portugal
    \and
    Faculdade de Ciências, Universidade do Porto, Rua do Campo Alegre, 4150-007 Porto, Portugal
    \and 
    IFPU -- Institute for Fundamental Physics of the Universe, via Beirut 2, I-34151 Trieste, Italy
    \and
    Hamburger Sternwarte, Universitaet Hamburg, Gojenbergsweg 112, D-21029 Hamburg, Germany
    \and
    European Southern Observatory (ESO), Karl-Schwarzschild-Str. 2, 85748 Garching bei Munchen, Germany
    \and
    INAF -- Arcetri Astrophysical Observatory, Largo E. Fermi 5, I-50125 Florence, Italy
    \and
    Instituto de Astrof\'{\i}sica de Canarias (IAC), Calle V\'{\i}a L\'actea s/n, E-38205 La Laguna, Tenerife, Spain\
    \and
    Departamento de Astrof\'{\i}sica, Universidad de La Laguna (ULL), E-38206 La Laguna, Tenerife, Spain
    \and
    Centre for Astrophysics and Supercomputing, Swinburne University of Technology, Hawthorn, Victoria 3122, Australia
    \and
    Instituto de Astrof\'isica e Ci\^encias do Espa\c{c}o, Faculdade de Ci\^encias da Universidade de Lisboa, Campo Grande, PT1749-016 Lisboa, Portugal
    \and
    Departamento de Física da Faculdade de Ciências da Universidade de Lisboa, Edifício C8, 1749-016 Lisboa, Portugal
    \and
    Observatoire Astronomique de l'Universit\'e de Gen\`eve, Chemin Pegasi 51, CH-1290 Versoix, Switzerland
    \and
    Center for Space and Habitability, University of Bern, Gesellschaftsstrasse 6, Switzerland
    \and
    Cerro Tololo Inter-American Observatory/NSF NOIRLab, Casilla 603, La Serena, Chile
    \and
    INAF -- Osservatorio Astronomico di Padova, Vicolo dell'Osservatorio 5, I-35122, Padova, Italy
    \and
    INAF – Osservatorio Astrofisico di Torino, Via Osservatorio 20, I-10025 Pino Torinese, Italy
    \and
    Centro de Astrobiología, CSIC-INTA, Camino Bajo del Castillo s/n, 28602 Villanueva de la Cañada, Madrid, Spain
    }

   \date{Received XXX; accepted YYY}

  \abstract
  {The measurement of the tiny temporal evolution in the redshift of distant objects, the redshift drift, is a powerful probe of universal expansion and cosmology.}
  {We perform the first steps towards a measurement of such effect using the Lyman-$\alpha$  forest in the spectra of bright quasars as a tracer of cosmological expansion. Our immediate goal is to determine to which precision a velocity shift measurement can be carried out with the signal-to-noise (S/N) level currently available and whether this precision aligns with previous theoretical expectations.
  A precise assessment of the achievable measurement precision is fundamental for estimating the time required to carry out the whole project. We also aim to study possible systematic effects of astrophysical or instrumental nature arising in the measurement.}
  {We acquire 12 hours of ESPRESSO observations distributed over $0.875$ years of the brightest quasar known, J052915.80-435152.0 ($z_{\rm em}=3.962$), to obtain high-resolution spectra of the Lyman-$\alpha$  forest, with median S/N of $\sim86$ per $1~\kms$ pixel at the continuum. 
  We divide the observations into two distinct epochs and analyse them using both a pixel-by-pixel method and a model-based approach. This comparison allows us to estimate the velocity shift between the epochs, as well as the velocity precision that can be achieved at this S/N. The model-based method is calibrated using high-resolution simulations of the intergalactic medium from the Sherwood Simulation Suite, and it provides greater accuracy compared to the pixel-by-pixel approach.}
  {We measure a velocity drift of the Lyman-$\alpha$  forest consistent with zero: 
  $\Delta v = -1.25_{ - 4.46} ^{+ 4.44}\; {\rm m\,s^{-1}}$, equivalent to a cosmological drift of $\dot{v}=-1.43 _{- 5.10} ^{+ 5.08} \; {\rm m\,s^{-1}\,yr^{-1}}$ or 
  $\dot{z}= -2.19_{- 7.78} ^{+ 7.75} \times 10^{-8}\;{\rm yr^{-1}}$.
  The measurement uncertainties are on par with the expected precision.
  We estimate that reaching a 99\% detection of the cosmic drift requires a monitoring campaign of 5400 hours of integration time over 54 years with an ELT and an ANDES-like high-resolution spectrograph. 
  }
  {}
  {}

\titlerunning{The ESPRESSO Redshift Drift Experiment I}
\authorrunning{A. Trost et al.}
   \maketitle

\section{Introduction}

Directly measuring the expansion history of the Universe is among the most pressing tasks of observational cosmology, especially considering the evidence for its recent acceleration phase 
\citep{Riess1998,Riess21,Garnavich98,DETF2006,Dawson13,Dawson16,ANDES,DESI2024}. Most studies aiming to do this include specific assumptions on an underlying model, or class of models, leading to cosmological constraints which are model-dependent. A conceptual alternative, not yet realised in practice but expected to be achieved by forthcoming astrophysical facilities, is the redshift drift measurement, also known as the Sandage test.

As first explored by \citet{Sandage62}, the evolution of the Hubble expansion causes the redshift of distant objects participating in the Hubble flow to change slowly with time. Just as the cosmological redshift $z$ provides evidence of the expansion, the drift in this redshift provides evidence of the expansion's acceleration or deceleration between the epoch $z$ and the present one. This implies that the expansion history of the Universe can be detected and mapped, at least in principle, by means of a straightforward spectroscopic monitoring campaign.

For simplicity, we take a generic metric theory of gravity with the further assumption of homogeneity and isotropy, in which case the evolution of the metric tensor is entirely specified by the global scale factor $a(t)$. 
A photon emitted by some object at comoving distance $\chi$ at the time $t_{em}$ and observed by us at $t_{obs}$, will have a redshift
\begin{equation}\label{eq:redshift}
    1+z(t_{obs},t_{em}) = \frac{a(t_{obs})}{a(t_{em})}.
\end{equation}
Now consider how the redshift of an object at a fixed comoving distance $\chi$ evolves with $t_{obs}$. Over a comparatively short timescale $\Delta t_{obs}$, it will drift by
\begin{equation}
    \frac{dz_\chi}{dt_{obs}} \approx \frac{z_\chi(t_{obs}+\Delta t_{obs})-z_\chi(t_{obs})}{\Delta t_{obs}}\,,
\end{equation}
which, differentiating Eq. \ref{eq:redshift} with respect to $t_{obs}$, leads to
\begin{equation}
 \frac{dz_\chi}{dt_{obs}}(t_{obs}) = \left[1+z_{\chi}(t_{obs})\right]H(t_{obs})-H(t_{em})\,.
\end{equation}
Finally, we obtain, in simplified form \citep{McVittie62}
\begin{equation}\label{eq:sandage}
    \dot{z}\equiv\frac{dz}{dt_{obs}}=(1+z)~ H_0-H(z).
\end{equation}
Measuring $\dot{z}$ for a number of objects at different $z$ gives the function $\dot{a}(z)$ and, given $a(z)$, it is possible to reconstruct $a(t)$. Therefore, a measurement of $\dot{z}$ amounts to a purely dynamical and model-independent reconstruction of the expansion history of the Universe.

The effect is conceptually simple and provides a tool for key cosmological tests. Inter alia, one can easily verify that in decelerating universes, one expects a negative redshift drift at all redshifts. Conversely, a positive drift signal implies a violation of the strong energy condition and hence the presence of some form of dark energy that accelerates the Universe \citep{Liske08,Uzan,Quercellini,Heinesen}.  Mapping the redshift evolution of the drift can also lead to constraints on parameters in specific models, either alone or in combination with traditional cosmological probes \citep{Alves:2019hrg,Rocha:2022gog,Marques1,Marques2}.

The theoretical framework outlined above highlights the immense potential of redshift drift measurements to directly probe the expansion history of the Universe in a cosmological model-independent manner. However, achieving the required sensitivity for detecting this effect is an observational challenge due to its exceedingly small expected magnitude. For instance, over a timescale of $\sim 10$ years, the redshift drift $\dot{z}$ in $\Lambda$CDM~\citep{Planck18} is typically of the order of $10^{-10}$ to $10^{-9}$, or few centimetres per second per decade in spectroscopic terms, for $0<z<6$.
Such precision necessitates both an adequate tracer population and a high-resolution spectroscopic capability that ensures the systematic uncertainties are smaller than the signal itself.

Among the possible tracers for this effect, the Lyman-$\alpha$ forest stands out as a particularly promising candidate \citep{Loeb98}. The Lyman-$\alpha$ forest consists of a series of absorption lines imprinted on the spectra of distant quasars due to intervening clouds of neutral hydrogen in the intergalactic medium (IGM).
The suitability of the Lyman-$\alpha$ forest for redshift drift measurements arises from several key factors:
\begin{itemize}
    \item Low cosmic overdensities: The Lyman-$\alpha$  forest arises from sparse intergalactic neutral hydrogen, which is physically unconnected with the background source, faithfully tracing the Hubble flow, apart from seldom strong absorbers due to dense intervening galactic environments \citep{Rauch98}.
    \item Abundance of tracers: The Lyman-$\alpha$ forest provides a dense sampling of absorption features along the line of sight (with number density $dn/dz\approx100-200$ at $z=2-3$, \cite{Kim2002}), enabling the measurement of the redshift drift across numerous independent systems at varying redshifts.
    \item Spectroscopic accessibility: Even though the Lyman-$\alpha$  lines are not particularly narrow, with a typical linewidth of $\sim 30~\kms$ \citep{Kim2002}, through the use of high-resolution spectroscopy their positions can be determined with sufficiently high precision to enable the measurement of the redshift drift. Moreover, the Lyman-$\alpha$  lines at $z>1.5$ are observable with large ground facilities.
    \item High-redshift sensitivity: At high redshifts, specifically deep in the matter era where the Universe is decelerating, the term $(1+z)$ in Eq.~\ref{eq:sandage} amplifies the magnitude of the drift, making the signal more detectable \citep[c.f. Fig. 2 in ][]{Liske08}.

\end{itemize}
Thus far, no detection of this signal is available, and existing upper limits based on other probes such as H~\textsc{i} 21cm absorption lines \citep{Darling2012} and narrow metal absorption lines \citep{Cooke2020},
have error bars three orders of magnitude larger than the expected signal. The first detections of the signal are expected to come from the ArmazoNes high Dispersion Echelle Spectrograph (ANDES) at the ESO's Extremely Large Telescope (ELT) \citep{Marconi2024, ANDES} and by the full configuration of the Square Kilometre Array (SKA) \citep{Klockner, Kang2025}, which are sensitive to the high redshift (matter era) and low redshift (acceleration era), respectively.

The Echelle SPectrograph for Rocky Exoplanets and Stable Spectroscopic Observations (ESPRESSO) spectrograph at the combined Coud\'e focus of ESO's VLT \citep{Pepe20}, has a similar design to ANDES and enables us to bridge the gap between the present and the future. As part of the ESPRESSO Science Team's Guaranteed Time Observations (GTO), we are carrying out the first dedicated redshift drift experiment, with a nominal experiment time of one year and an observation time of four nights (ca. 40 hours). This paper presents the first results, derived from 12h of observations of one quasar target.

Our two overarching goals are to improve the aforementioned upper limits and to start developing, testing, and validating a full end-to-end analysis pipeline, which will provide a baseline for the corresponding pipeline for the ANDES redshift drift measurement. This is the first of a series of reports on this experiment.

Specifically, in this work we set the first two epochs of observations with high-resolution, high-stability ESPRESSO spectroscopy and estimate the achievable precision. We check whether this aligns with previous estimates and if systematic effects of astrophysical or instrumental nature appear at the present level of signal-to-noise (S/N). We extrapolate our estimates to a realistic ANDES observational campaign to estimate the required amount of time to achieve a significant detection of the cosmological expansion. 

In Sect.~\ref{sect:target} we present the quasar targeted for our study. In Sect.~\ref{sect:data} we describe the observation taken with ESPRESSO, the data reduction, the definition of the two observational epochs and the recognition of metal lines and strong absorbers throughout the spectrum. The expected precision of a velocity drift measurement at the current level of S/N is presented in Sect.~\ref{sect:liske_pred}. In Sect.\ref{sect:direct} we apply a direct pixel-by-pixel method, based on the work of \citep{Bouchy2001}, to estimate the velocity shift $\Delta v$ that has occurred between the two epochs. A different method to perform the measurement, based on modelling the Lyman-$\alpha$  forest, is presented in Sect.~\ref{sec:model_based}. In Sect.~\ref{sec:validation} we investigate the validity of our estimates and tackle the possible systematic effects arising in the measurement due to astrophysical or instrumental effects. In Sect.~\ref{sec:future} we extrapolate our results to estimate the total observational time required to carry out a significant detection of the cosmological drift, either with ESPRESSO or with ANDES. Finally, we summarise and discuss our results in Sect.~\ref{sec:discussion}.

\section{Target selection}\label{sect:target}
The QUBRICS (QUasars as BRIght beacons for Cosmology in the Southern hemisphere) survey \citep{Qubrics23} 
provides a Golden Sample of seven bright quasars specifically selected to carry out the measurement of the redshift drift, observing their Lyman-$\alpha$  forest with an ESPRESSO- or ANDES-like spectrograph. For the present case study and the first epoch observation, we consider the brightest object from the Golden Sample, J052915.80-435152.0, hereafter called Super Bright 2 (SB2). In a parallel paper (Marques et al. in prep.) we will discuss the observation of the second-most luminous object of the sample J212540.97-171951.4 (hereafter called SB1)\footnote{The terminology "Super Bright quasars" used in this paper is purely chronological and does not reflect hierarchy based on luminosity, originating from the observation sequence: J212540.97 was identified at an earlier stage in the QUBRICS survey and is designated as SB1, whereas the brighter J052915.80 was discovered later.}.
SB2 has the identifier QID 1128023 in the QUBRICS database \citep{Calderone2019} with coordinates (J2000) RA 05:29:15.81 and Dec -43:51:52.1 and magnitudes $r=16.3005$, $i=16.1184$ \citep{Skymapper4} and $G=16.3452$ \citep{GAIADR3}

Besides being particularly suited for the Sandage test, SB2 is an outstanding object, the most luminous quasar known in the Universe. Medium resolution spectra taken with X-Shooter \citep{Vernet2011} have revealed that SB2 has a bolometric luminosity of $\log(L_{\rm bol}/{\rm erg\; s^{-1}}) = 48.27$, supported by an accretion of $370 M_\odot$ per year onto a supermassive black hole (SMBH) of mass $\log M/M_\odot\sim10.24$ or, in other terms, about 1 solar mass per day accreting onto a SMBH of $\sim17$ billion solar masses. From the same data, \cite{Wolf2024} report an emission redshift for SB2 of $z_{em}=3.962$. 

\section{Data acquisition and treatment}\label{sect:data}
In this work, we present high-resolution, high S/N optical spectroscopy of this outstanding object
obtained with ESPRESSO.
ESPRESSO observations have been carried out using the single UT mode \citep{Pepe20} and achieve a resolving power of $R=\lambda/\Delta\lambda\sim135000$, where $\Delta\lambda$ is the full width at half maximum of resolution element ($\sim2.2 ~\kms$ in velocity space), with a 4x2 detector binning, chosen to optimize the S/N in the region covering the Lyman-$\alpha$  forest.
A summary of the observations is given in Table~\ref{tab:observations}. 

\begin{table*}
    \centering
        \caption{Summary of single Observing Blocks (OB) of SB2 taken during observing periods P110 and P112.  
}
    \begin{tabular}{lcccccc}
    \hline
        Period & Exposure Start (UTC) & MJD & $t_{exp}$ & Grade & Seeing & Airmass   \\
               &                && [s]         &       & [arcsec] &   \\
    \hline
     P110 & 2022-10-29 07:21:01 & 59881.31068001& 3438.0 & C & 0.93 & 1.066 \\
          & 2023-01-22 01:04:40 & 59966.04843679& 3438.0 & A & 0.62 & 1.071 \\
          & 2023-01-22 02:13:07 & 59966.09453567& 3438.0 & B & 0.72 & 1.075 \\
          & 2023-01-23 01:14:56 & 59997.04166665& 3438.0 & C & 1.28 & 1.066 \\
          & 2023-02-22 00:54.53 & 59967.05436154& 3438.0 & A & 0.60 & 1.116\\ \hline
     P112 & 2023-08-25 08:41:12 & 60285.19820316& 3438.0 & A & 0.42 & 1.278 \\
          & 2023-11-11 05:22:38 & 60284.18013713& 3453.0 & C & 0.99 & 1.087 \\
          & 2023-11-19 05:02:35 & 60345.18147687& 3453.0 & A & 0.64 & 1.078 \\
          & 2023-12-06 04:15:02 & 60259.22743581& 3453.0 & A & 0.49 & 1.069 \\
          & 2023-12-07 04:41:30 & 60267.21411237& 3453.0 & A & 0.56 & 1.065 \\   
          & 2023-12-11 04:47:35 & 60289.20252708& 3453.0 & A & 0.57 & 1.070 \\
          & 2023-12-13 05:22:56 & 60291.22794520& 3453.0 & A & 0.54 & 1.103 \\
          & 2024-02-05 04:15:20 & 60181.36712803& 3453.0 & C & 0.50 & 1.512 \\
          
    \hline

    \end{tabular}
    \label{tab:observations}
    \tablefoot{The first column specifies the ESO period in which the observations were collected. The second column gives the date and hour of the start of observations (UT), which is also reported in MJD format in the third column. The fourth column gives the exposure time in seconds. The fifth column reports the OB grade, given by ESO on a set of constraints on observing parameters (image quality, airmass, sky transparency, fractional moon illumination, moon distance, precipitable water vapour, twilight allowance): `A' means that all constraints are fulfilled; `B' and `C' mean that at least one constraint is exceeded by less than 10\% or more than 10\% respectively. The sixth and seventh columns report the seeing and airmass, respectively, as quoted from the FITS files' headers. Upon inspection, grade B and C observations did not show a decrease in the achieved S/N. Exceptionally, we consider them for our purposes since we do not expect any strong systematic influence on the final results at our level of S/N. In the future, only grade A observations will be considered.}
\end{table*}

The observations have been reduced using the ESPRESSO Data Reduction Software\footnote{\url{https://www.eso.org/sci/software/pipelines/espresso/}} (DRS, \citealt{DiMarcantonio2018}), version 3.1.0. 
Laser Frequency Combs (LFC) frames were used for wavelength calibration of the spectra. 
We also produced spectra calibrated on the Fabry-P{\'e}rot (FP) etalon combined with a ThAr lamp frames to investigate the presence of possible instrumental systematics due to calibration stability and accuracy (see Sect.~\ref{sec:instrumental_systematics}).

The final products of the pipeline were flat-fielded, blaze-corrected, sky-subtracted and wavelength-calibrated.
The spectra were also corrected to the Solar System barycentre using standard pipeline procedures, employing INPOP13c ephemeris \citep{INPOP13}. Earth's precession is taken into account in the correction, consistently with the requirements for achieving sub-${\rm m\, s^{-1}}$ precision \citep{WrightEastman2014}.

and barycentric-corrected spectra, both in order-by-order and order-merged format. 

Order-by-order spectra were optimally extracted along detector rows (in the cross-dispersion direction), retaining for each pixel the original calibrated wavelength of the rows themselves, while the order-merged spectra were rebinned to a common wavelength grid.

The coaddition of the extracted spectra was performed using the \textsc{Astrocook} software\footnote{\url{https://github.com/DAS-OATs/astrocook}} \citep{Cupani20}, following the `drizzling'-like approach described in \citet{Cupani16}. A wavelength grid was defined with a fixed step of $1~\kms$, and the pixels contributing to each bin in the grid were selected from the order-by-order spectra, to combine them all together in a single rebinning procedure. Contributing pixels were weighted by their associated error and by how much they overlapped with the wavelength range of the bins themselves. Order-by-order spectra were equalised prior to coaddition by multiplying their flux by scalars to get the same median flux level, to eliminate differences in flux levels due to varying observing conditions.

\begin{figure*}
    \centering
    \includegraphics[width=\linewidth]{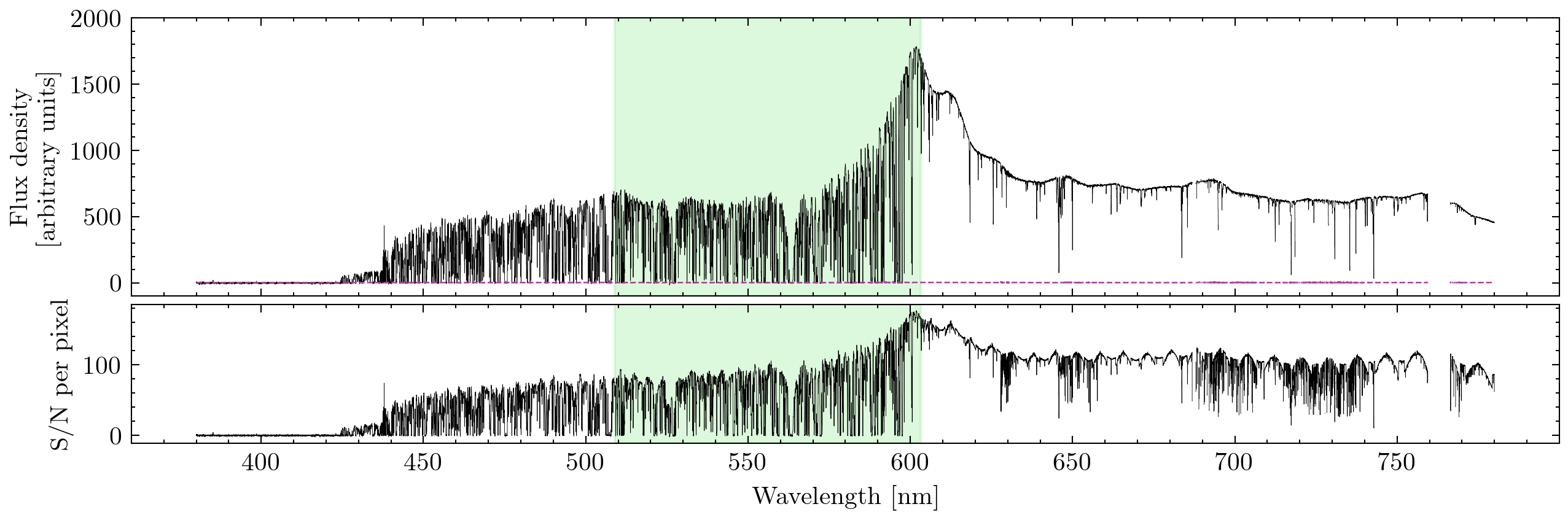}
    \caption{Combined spectrum of SB2. Top panel: Flux density in arbitrary units is shown in black, with the dashed purple line denoting the flux density error. Bottom panel: S/N per $1~\kms$ pixel. The green shaded area highlights the Lyman-$\alpha$  forest considered in the redshift drift measurement, bound by the Lyman-$\alpha$  and Lyman-$\beta$  emissions of the quasar, namely between $509 - 603~\unit{\nano\meter}$.}
    \label{fig:spectrum}
\end{figure*}

Coaddition of individual exposures resulted in a single spectrum, with a median S/N~$\sim86$ per $1~\kms$ pixel at the continuum level in the forest, and $\sim105$ redward of the Lyman-$\alpha$  emission, as shown in Fig.~\ref{fig:spectrum}.
The Lyman-$\alpha$  forest highlighted in the figure is defined as the spectral interval between the Lyman-$\alpha$  and Lyman-$\beta$  emission of the quasar, namely $509 - 603~\unit{\nano\meter}$, shown in more detail in Fig.~\ref{fig:forest_plot}. This section of the spectrum contains Lyman-$\alpha$  absorption lines with $3.18\le z\le3.96$, and no Lyman-$\beta$  absorptions.

Observations taken in the two ESO observing periods P110 and P112 (see Table~\ref{tab:observations}), respectively, were also combined separately to provide two independent spectra that were used as the two epochs of the Sandage test. The 1st (P110) and 2nd (P112) epoch spectra have a median S/N level at the continuum in the Lyman-$\alpha$  forest of $\sim 47$ and $\sim72$, respectively. We defined the temporal baseline occurring between the two epochs' spectra as the difference between the mean observational date of each period $\Delta t = 0.875 ~{\rm yr}$. 

\subsection{Removing metal transitions from the Lyman-$\alpha$  forest}
As shown by \cite{Cristiani24}, Lyman-$\alpha$  and metal lines have significantly different dynamical behaviour on the temporal and physical scales probed by the experiment. The authors showed that absorption lines due to neutral hydrogen on adjacent sightlines (with sub-kpc separations) appear identical within the noise, whereas metal lines show significant differences in their velocity structure. These intrinsic dynamical effects can be a source of strong systematic effects, inducing a velocity shift in the metal lines up to a few $\unit{\cmsyr}$, of the same order of magnitude as the expected cosmic signal. We therefore removed the metal-polluted regions from our analysis to reduce possible systematics. 

This process was done using the \textsc{Astrocook} software \citep{Cupani20} to provide a secure identification of metal lines in the spectrum. First, the quasar continuum emission was estimated by applying an iterative sigma-clipping procedure to remove absorption
features while masking sky emission lines, telluric absorption lines, or spurious, narrow spikes in flux (e.g. `cosmic rays' that were not removed by the coadding process), and manually correcting the continuum level where necessary.
Sky and telluric lines were masked using a reference {\sc Skycorr} model \citep{2014A&A...567A..25N} and empirically defining proper cuts to conservatively remove even the line wings. 
We performed a search for metal absorption lines throughout the spectrum, starting from line doublets (e.g. C~\textsc{iv}, Si~\textsc{iv} and Mg~\textsc{ii}) redward of the Lyman-$\alpha$  emission and following up with associated lines, i.e. other ionic transitions at the same redshift as the found doublets. Table~\ref{tab:metalsysts} reports the absorption systems, i.e. the groups of ionic transitions with the same redshift, found along the spectrum of SB2.

\begin{table}
    \centering
        \caption{Metallic absorption systems identified in the spectrum of SB2.}
    \begin{tabular}{lr}
    \hline
        $z$ & Ions \\
        \hline
         0.96 &                           Fe \textsc{ii}, Mg \textsc{i}, Mg \textsc{ii}\\
         1.13 &                           Fe \textsc{ii}, Mg \textsc{i}, Mg \textsc{ii}\\
         1.44 &                           Fe \textsc{ii}, Mg \textsc{ii}\\
         1.63 &                           Fe \textsc{ii}, Mg \textsc{ii}\\
         1.84 &                           Al \textsc{ii}, Al \textsc{iii}, Fe\textsc{ii}\\
         2.12 &                           Al \textsc{ii}, Fe \textsc{ii},  Si\textsc{ii}\\
         2.30 &                           Al \textsc{ii}, Al \textsc{iii}, C \textsc{ii}, C \textsc{iv}, Si \textsc{ii}, Si \textsc{iv}\\
         2.33 &                           C \textsc{iv}\\
         2.46 &                           C \textsc{iv}, Si \textsc{iv}\\
         2.91 &                           C \textsc{iv}\\
         3.01 &                           C \textsc{iv}\\
         3.13 &                           C \textsc{iv}, Si \textsc{iii}\\
         3.17 &                           Al \textsc{ii}, C \textsc{ii}, C \textsc{iv}, Si \textsc{ii}, Si \textsc{iii}, Si \textsc{iv}\\
         3.29 &                           C \textsc{iv}\\
         3.33 &                           Al \textsc{ii}, C \textsc{ii}, C \textsc{iv}, O \textsc{i}, Si \textsc{ii}, Si \textsc{iv} \\
         3.29 &                           C \textsc{iv}\\
         3.49 &                           C \textsc{iii}, C \textsc{iv}, Si \textsc{iii}, Si \textsc{iv}\\
         3.59 &                           C \textsc{iv}\\
         3.60 &                           C \textsc{iii}, C \textsc{iv}, Si \textsc{iii}, Si \textsc{iv}\\
         3.63 &                           Al \textsc{ii}, C \textsc{ii}, C \textsc{iv}, Fe \textsc{ii}, O \textsc{i}, S \textsc{iv}, Si \textsc{ii}, Si \textsc{iv}\\
         3.67 &                           C \textsc{iv}, Fe \textsc{iii}\\
         3.71 &                           C \textsc{iv}, Si \textsc{iv}\\
         3.78 &                           C \textsc{iv}, Si \textsc{iii}, Si \textsc{iv}\\
         3.80 &                           C \textsc{iv}, Si \textsc{iii}, Si \textsc{iv}\\
         3.84 &                           C \textsc{iii}, C \textsc{iv}, Si \textsc{iv}\\
         3.86 &                           C \textsc{iv}\\
         3.90 &                           C \textsc{iii}, C \textsc{iv}\\
    \hline
    \end{tabular}
    \label{tab:metalsysts}
    \tablefoot{The left column reports the system redshift, and the right column the identified atomic species.}
\end{table}
Once identified, we fit composite Voigt profiles to the metal absorption lines that fall redward of the Lyman-$\alpha$  emission, estimating the absorber's redshift, column density, and Doppler broadening. Metal lines blueward of the Lyman-$\alpha$  emission were not fitted, as this would require modelling multiple blended H~\textsc{i} components, making the analysis significantly more complex. Since this is beyond the scope of the present study, we left it unaddressed and masked out such lines, removing the spectral regions in velocity space associated with higher wavelength transitions.
Table~\ref{tab:metal_list} reports the fitted Voigt parameters for the metal transitions found redward of the Lyman-$\alpha$  emission. 

\subsection{A sub-DLA at $z_{\rm abs}=3.63$}
As visible in Fig.~\ref{fig:spectrum}, the spectrum of SB2 presents a strong H~\textsc{i} absorption at $\lambda\sim 5 63~\unit{\nano\meter}$ or $z\sim3.63$. 
We study this absorbing system in more detail to understand, based on its column density, whether it might belong to a dense environment prone to proper motions related to galactic dynamical processes, or it stems from the sparse cosmic web following the Hubble flow and thus can be considered in the redshift drift measurement. 
We found strong and complex associated metal absorption redward of the Lyman-$\alpha$  emission, stemming from ions with both low (O~\textsc{i}, Al~\textsc{ii}, C~\textsc{ii}, Fe~\textsc{ii}, Si~\textsc{ii}) and high (C~\textsc{iv}, Si~\textsc{iv}) ionization states.
We fitted the Lyman-$\alpha$  absorber within Astrocook using 3 components, tying the redshifts of H~\textsc{i} and low ionisation species. The total H~\textsc{i} column density of the system is $\log N_{H~\textsc{i}}\approx19.8$, corresponding to a sub-Damped Lyman-$\alpha$  System (sub-DLA, \cite{Peroux2003}).

\begin{figure*}
    \centering
    \includegraphics[width=\linewidth]{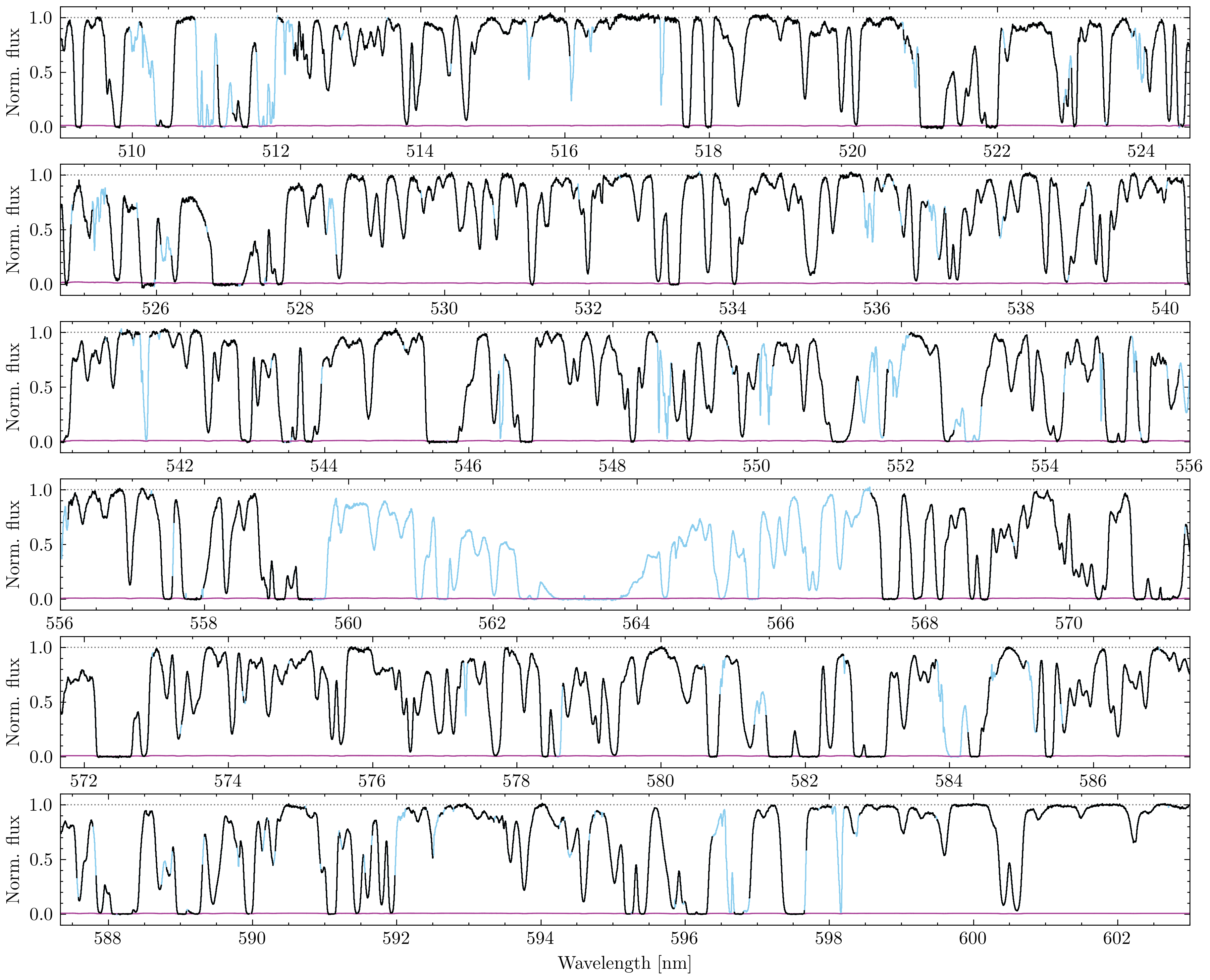}
    \caption{Normalised transmitted flux of the Lyman-$\alpha$  forest in the combined spectrum of SB2 as a function of wavelength. The light blue solid line highlights the spectral regions masked due to metal absorption, the sub-DLA and bad pixels. The purple solid line reports the flux error.}
    \label{fig:forest_plot}
\end{figure*}

Given this value, we decided to exclude the spectral range tied to this absorber from the analysis, as it is probably related to an intervening galactic system and could potentially induce systematics in the measurement due to its peculiar motions. We thus mask the region of the spectrum spanning $\pm 2000 ~\kms$ from the column density weighted average redshift position of the system. 
The final masked spectra used in our analysis have 41737 pixels ($1~\kms$ wide) in the Lyman-$\alpha$  forest, compared to 50805 pixels without mask. Meaning we consider only $82.15\%$ of the total Lyman-$\alpha$  forest.   
Fig.~\ref{fig:forest_plot} shows the normalised flux of the Lyman-$\alpha$  forest in the spectrum of SB2, highlighting the masked-out regions.

\section{Expected velocity precision}\label{sect:liske_pred}
In a $\Lambda$CDM~Universe \citep{Planck18}, the expected cosmological redshift drift at redshift $z=3.573$ (i.e. the middle of SB2's Lyman-$\alpha$  forest) is $\dot{z}=-6.5\times10^{-11} ~{\rm yr^{-1}}$, equivalent to $\dot{v}=-0.43 ~\unit{\cmsyr}$ in velocity space. On the baseline of our experiment, the expected velocity shift between the two epoch spectra is therefore $\Delta v = -0.38~\cms$.

\cite{Liske08} provided a practical formula to assess the precision expected from an experiment performed on the ELT based on an ensemble of quasars at different redshifts. We adapted this formula to estimate the precision expected from our data. 
Given the SB2 ESPRESSO spectra at the two epochs, we estimated the predicted precision in the velocity shift measurement, adapting the scaling relation of \cite{Liske08}
\begin{equation}\label{eq:vel_uncertainty_liske}
    \sigma_v = g\times 1.35 \left(\frac{\textrm{S/N}}{4075}\right)^{-1}\left(\frac{N_{\rm QSO}}{30}\right)^{-0.5}\left(\frac{1+z_{\rm QSO}}{5}\right)^{-\gamma} f_{\rm Ly_{\alpha}} ^{-0.5} \text{cm s}^{-1}
\end{equation}
where ${\rm S/N=86}$ is the median signal-to-noise ratio at the continuum level per $1~\kms$ pixel per object, accumulated over all observations (i.e. the S/N level of the combined spectrum). $N_{\rm QSO}$ is the number of quasars in the sample ($N_{\rm QSO}=1$ in our case), and $z_{\rm QSO}=3.962$ is the redshift of the quasar.
The $\gamma$ exponent is $1.7$ for $z_{\rm QSO}\le 4$ and $0.9$ above.
The factor $f_{\rm Ly_{\alpha}}=0.8215$ is the fraction of the Lyman-$\alpha$  forest considered after masking. 
The factor $g$, called {\it form factor}, depends on the observing strategy,
being 1 if half of the exposures are taken at the beginning of the experiment and half at the end, and becoming larger if the measurements are spread out over time, reaching 1.7 for a uniform distribution (see Sect.~6 of \cite{Liske08} for further details on the computation of the form factor). In our case, given the fact that 38\% of the total integration time is collected at the 1st epoch, and 62\% at the 2nd, the form factor has a value of $g=1.03$.
From Eq.~\ref{eq:vel_uncertainty_liske}, we can estimate the expected maximum precision allowed by photon noise on our current data to be $\sigma_v = 4.02 ~\ms$.

Note that in this work we analysed only one sightline, whereas the relation from which Eq.~\ref{eq:vel_uncertainty_liske} is derived has been calibrated on a measurement based on an ensemble of spectra. Sightline-to-sightline deviations from the predicted measurement precision due to cosmic variance are expected due to the varying number of Lyman-$\alpha$  lines and strong absorbers found in the spectra of different objects. 

\section{Redshift drift measurement: pixel-by-pixel method}\label{sect:direct}

We first measured $\Delta v$ using a method developed for measuring radial velocity shifts between repeated observations of stars for exoplanet search \citep{Bouchy2001}, which we refer to as the pixel-by-pixel method.
In summary, due to the drift of the absorption lines, the flux of the \textit{i}-th pixel at the second epoch, $F_{2,i}$,  can be expressed as a small perturbation of the flux in the same pixel at the first epoch, $F_{1,i}$:
\begin{equation}\label{eq:gradiente_method}
    F_{2,i} = F_{1,i} + \frac{dF_i}{d\lambda}\frac{\delta v_i}{c}\lambda_i,
\end{equation}
where $\lambda_i$ is the observed wavelength of the \textit{i}-th pixel, and $dF_i/d\lambda$ is the slope of the spectrum at that pixel. The perturbation of the fluxes between two epochs defines, for each pixel, a small velocity shift $\delta v_i$. Note that Eq.~\ref{eq:gradiente_method} holds only for $\delta v_i$ much smaller than the typical line width ($\sim 30~\kms$).
The single pixel's velocities can be averaged over the whole spectrum to estimate the velocity drift occurring between the two epochs. 
\begin{equation}
    \Delta v = \frac{\sum \delta v_i w_i}{\sum_i w_i},
\end{equation}
where each pixel is weighted by $w_i=\sigma^{-2}_{v_i}$, i.e. the inverse variance of $\delta v_i$. Accounting for the uncertainties in both epochs' spectra and in their derivative, the variance becomes
\begin{equation}
    \sigma^2_{v_i} = \left[ \frac{c}{\lambda_i (dF_i/d\lambda)}\right]^2\left[ \sigma_{1,i}^2+\sigma_{2,i}^2+ \frac{(F_{2,i} - F_{1,i})^2}{(dF_{i}/d\lambda)^2}\sigma^2_{F'_i}\right],
\end{equation}
where $\sigma_{1,i}$ and $\sigma_{2,i}$ are the \textit{i}-th pixel flux error in the two spectra, and $\sigma_{F'_i}$ is the error on the flux slope at the \textit{i}-th pixel. With this choice of weights, the uncertainty on the final $\Delta v$ estimate is 
\begin{equation}
    \sigma_v = \left[\sum_i \sigma^{-2}_{v_i}\right]^{-1/2}.
\end{equation}

It is worth noting that, at first order, the flux slope term $dF_i/d\lambda$ is epoch-invariant (under the assumption of very small shifts) and does not carry any epoch-dependent information. 
For this reason, we computed the flux derivative by means of finite differences of non-adjacent points, on the second epoch spectrum to exploit the higher S/N level and lower the noise in its evaluation.

We evaluated the velocity shift $\delta v$ occurring between the two epochs, exploiting the normalised fluxes in the Lyman-$\alpha$  forest of the two spectra, where we have masked metal lines, the strong H~\textsc{i} absorber at $z\sim3.63$, and bad pixels. 
Our estimate yields $\Delta v = -2.67 \pm 6.64 ~\ms$. Taking into account the mean baseline between the two epochs of $\Delta t = 0.875$ years, we estimate a drift of $\dot{v} = -3.06 \pm 7.59 ~\ms{\rm yr^{-1 }}$, equivalent to a redshift drift of $\dot{z} = (-0.47 \pm 1.15) \times 10^{-7}~{\rm yr^{-1 }}$

\section{Redshift drift measurement: model-based}\label{sec:model_based}
The larger than expected uncertainty from the direct method is surprising and is not completely understood whether it stems from either instrumental systematics, problems in the measurement procedure or it is intrinsic to our sightline. 
Loss of precision prolongs the time required for reliably detecting the redshift drift signal and raises the risk of critical failure. A secure estimate of the measurement uncertainty is therefore of crucial impact on the experiment timeline.

In this section, we developed a different approach to the measurement to investigate whether a higher precision can be achieved with the current data. 
Instead of directly comparing the two epochs' spectra, we built an ensemble of analytical models of the forest of SB2 that are then correlated to the two epochs. The modelling procedure is calibrated on mock spectra extracted from state-of-the-art physical simulations of the IGM and is thus `informed' on the spectral scales of the single Lyman-$\alpha$  lines and is able to de-prioritise pixels that might be affected by metal transitions or outlier pixels (e.g. non-removed cosmic rays). With this approach, we are capable of reducing the uncertainty in the measurement by a significant fraction and decreasing the total amount of integration time needed for the experiment. 

Summarising, the main steps of the method are:
\begin{itemize}
    \item Production of realistic mock sightlines that reproduce the properties of the ESPRESSO spectra of SB2 considered in the measurement.
    \item Calibration of the modelling procedure on the mock data.
    \item Construction of an ensemble of models of the Lyman-$\alpha$ forest of SB2.
    \item Evaluation of the velocity shift between the model ensemble and the 1st and 2nd epoch data.
\end{itemize}

We stress that the mock data were only used to calibrate the modelling procedure and are not involved in the final measurement. 

\subsection{Mock spectra}
We used realistic mock spectra that reproduce the properties of the three SB2 spectra (1st epoch, 2nd epoch, and combined) to calibrate the modelling pipeline and validate the measurement results. 
Starting from the hydrodynamical cosmological simulations of the IGM of the Sherwood Simulation Suite \citep{Bolton17} (with box size of $10$ Mpc/h and $2\times 1024^3$ dark matter and gas particles), we extracted the gas particle data and computed the H~\textsc{i} optical depth along 5000 skewers piercing the box at every redshift step $\Delta z=0.1$, within $3.1\le z \le 4$. 
The optical depth of the spectra was rescaled to match the mean optical depth observed at $z\sim3.57$ (centre of the forest) from \cite{Becker13}. The sightlines ($10$ Mpc/h long) are randomly selected from the appropriate redshift interval, shifted so that they begin and end with regions of no absorption, and stitched together to form 100 spectra that match the length of the Lyman-$\alpha$  of SB2 ($\sim421$ Mpc/h long, or $50321~\kms$). 

The long spectra were then convolved with a Gaussian of FWHM $\sim2.2~\kms$ to mimic ESPRESSO’s resolution and re-binned to a grid of pixels $1~\kms$ wide (the average native pixel size extracted from the simulation was $0.593~\kms$), matching the binning of our data. 
The noise properties of the original spectrum were added to the mocks following the prescription of \citet{Rorai2017} \citep[see also][]{Trost2024}. Summarising, the original 1st epoch spectrum was divided into 10 chunks, $\sim10$ nm wide. For each of these chunks, we:
\begin{itemize}
    \item divided the pixels into 50 bins according to their normalised flux value, producing 50 corresponding noise distributions; 
    \item assigned a noise value to each simulated pixel, where the value was randomly sampled from the distribution associated with the pixel's flux value, as defined in the point above.
\end{itemize}

This procedure was repeated for each chunk of the 1st epoch spectrum and its corresponding mock spectra, to preserve the relation between S/N at the continuum level and wavelength. 

The second epoch spectra were generated in the same way from the same sightlines. However, before sightline extraction, we artificially added a velocity drift of $-0.38 ~\cms$, i.e. the expected drift at $z\sim3.57$ for a $\Lambda$CDM~Universe during our baseline $\Delta t = 0.875 \,{\rm yr}$, to the simulated particles’ velocities. These sightlines were then processed in the same way, and the noise of the 2nd epoch was added to the flux, as explained above.

\begin{figure}
    \centering
    \includegraphics[width=1\linewidth]{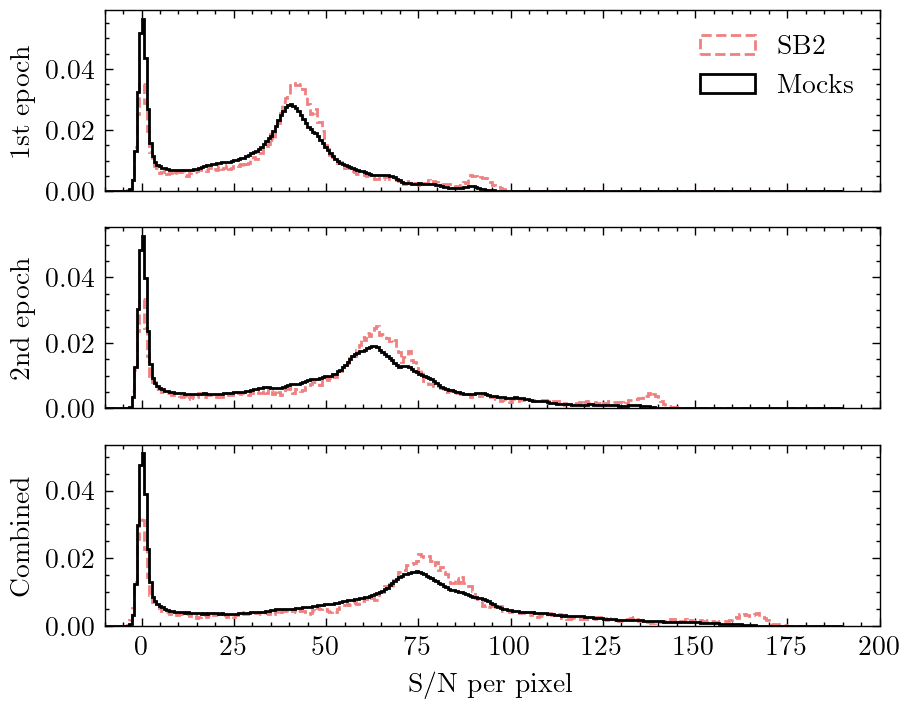}
    \caption{Distribution of the S/N per pixel in the Lyman-$\alpha$  forest of the three spectra of SB2 (dashed red) and of the corresponding mock spectra (solid black). Upper panel: first epoch, middle panel: second epoch, bottom panel: combined spectrum.}
    \label{fig:snr}
\end{figure}

We combined the first and second epochs' mock datasets via a weighted average, with weights proportional to the inverse of the pixels' flux variance, to create the mock combined spectrum. 
Fig.~\ref{fig:snr} shows the distribution of the S/N per pixel in the three mock datasets and the original spectra.
To further validate the analysis (see Sect.~\ref{sec:validation}), we also built second epoch mock datasets assuming either no drift between the two epochs or $10^3$ and $10^4$ years of temporal separation, adding a velocity drift to the simulation's particles of $-3.8~\ms$ and $-37.6~\ms$, respectively. 

On all spectra, we applied the same pixel masking of the original data, derived from metal, sub-DLA and bad pixels' removal, to ensure the absence of systematics in the final measurement due to pixel masking and to conserve the total amount of pixels involved in the measurement.

\subsection{Model assessment}
We relied on mock data to calibrate the modelling procedure before building the best possible model of SB2. Multiple sightlines were considered in the calibration to properly account for cosmic variance effects in the modelling procedure.
Given the short time-scales between the two epochs and that the total S/N is insufficient to detect the cosmic drift signal, we simplified the procedure by creating a fiducial model of the combined spectrum instead of modelling the two epoch spectra individually. A further simplification regards our decision to model the spectrum using cubic splines\footnote{Using higher order splines does not affect the results of the analysis.} instead of the more traditional approach using Voigt profile decomposition. This was done because we were not interested in measuring physical properties of the gas but rather differences between the SB2's spectra in the two epochs, for which the spline method was sufficient.

In quasar spectra, Lyman-$\alpha$  lines have a typical width of $\sim 30~\kms$ \citep{Kim2002}, while outlier features (e.g. cosmic rays) and metal transitions are typically narrower (down to few $\unit{\kms}$). 
We calibrated the spline knot spacing on mock spectra containing only Lyman-$\alpha$  absorption, so that our procedure was `informed' on the scales that we want to probe. Then, by generating an ensemble of models for SB2, we recovered a higher variance among models in the regions that might be affected by sharp non-Lyman-$\alpha$  absorption and could de-prioritise such features in the final measurement of $\Delta v$. The described procedure also prevents over-fitting of the noise.

In practice, let $S^k(\lambda|{\bf n}(A, \phi))$ be the spline model fitted on the \textit{k}-th spectrum of the combined mock dataset described in the previous section, using the square inverse of the flux noise as weights in the fit, where ${\bf n}$ is a vector defining the positions of the spline knots along the spectrum. The knots are equispaced in velocity space, separated by a constant velocity $A$, starting from an initial position (or phase) $0<\phi<A$ with respect to the beginning of the Lyman-$\alpha$  forest. 

Goodness of fit for the model was determined from its residuals to the data, $R^k(\lambda|A,\phi) = \left(F - S^k(\lambda|{\bf n}(A, \phi))\right)/\sigma_F$. We considered a model to be good if the residuals are centred at zero with a variance equal to unity, $R^k\sim\mathcal{N}(0,1)$. 

\begin{figure}
    \centering
    \includegraphics[width=1\linewidth]{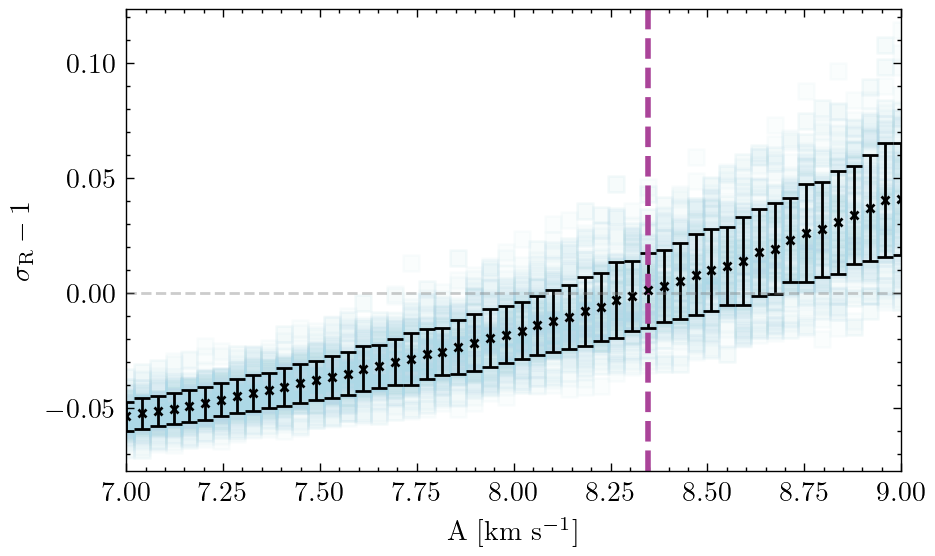}
    \caption{Standard deviation of the fit residuals between the combined mock spectra's flux and the spline model with internodal distance $A$. The blue squares describe the $\sigma_R$ of models on different mock spectra and randomly drawn initial node phase $\phi$. The black error bars define the average and standard deviation of $\sigma_R(A)-1$. The vertical purple dashed line denotes the optimal value of $A$.}
    \label{fig:Residuals-sigma}
\end{figure}

One hundred mock spectra were fitted in this way, varying both the inter-nodal distance $A$ and the initial phase $\phi$, and their residuals were calculated as above. For each value of $A$ and $\phi$, we calculated the mean and the variance of the residuals for all 100 mock spectra. The optimal value of $A$ was determined by examining which combination of parameters results in residuals most closely matching values drawn from $\mathcal{N}(0,1)$. Mathematically, we searched for a value of $A$ that minimises:
\begin{equation}\label{eq:chi_A}
    \varepsilon(A) = \frac{\left(\sigma_{\rm R}(A)-1\right)^2}{\sigma^2_{\sigma_{\rm R}}(A)},
\end{equation} 
where $\sigma_R(A)$ is the standard deviation of the fit residuals of splines with internodal distance $A$, averaged over all spectra and all considered initial node phases. $\sigma^2_{\sigma_{\rm R}}(A)$ is the variance of the quantity $\sigma_R(A)$ due to sightline variance and variations of the initial node phase. Fig. \ref{fig:Residuals-sigma} shows $\sigma_R-1$ and its variance as a function of internodal distance $A$. Minimizing Eq.~\ref{eq:chi_A} yields a best inter-nodal distance of $A = 8.37^{+0.26}_{-0.24}~\kms$.
Note that this value of $A$ is valid only for this precise S/N level, resolution, pixel size and redshift. For any change in the spectral properties, $A$ should be re-evaluated.

\subsection{Building the forest model}\label{sec:model_build}

Given the optimal knot spacing and its uncertainty, $A$ and $\sigma_A$, we were able to construct a large number of models of the Lyman-$\alpha$  forest of SB2 that describe the data equally well in terms of a goodness-of-fit statistic (e.g. $\chi^2$). This allowed us to more thoroughly assess the impact of model non-uniqueness on the final result. This is conceptually similar to the approach taken by, for example, \citet{Lee2021MNRAS.507...27L,Milakovic2024MNRAS.534...12M,Webb2025MNRAS.tmpL..11W} in the context of fundamental constant and isotopic ratio measurements.  Additionally, we identified parts of the spectrum where model non-uniqueness was minimal, such that those regions can be prioritised in our analysis, whereas regions where model diversity (and hence uncertainty) was larger are considered `less trustworthy'. Moreover, we implement a weighting scheme based on the model's gradient and its uncertainty, mimicking the pixel-by-pixel method (Sect. \ref{sect:direct}), to achieve a larger sensitivity to velocity shifts. Both aspects are major benefits of the model-based method, which results in a better handling of uncertainties. 

We started by fitting the combined SB2 spectrum with $N=500$ splines $S_j(\lambda\,|\,{\bf n}_j)$. The node positions of the \textit{j}-th model, ${\bf n}_j$, were generated starting from an equispaced sequence of nodes in velocity space ${\bf n}_{j,0}$, with inter-nodal separation $A_j$ drawn from a normal distribution $\mathcal{N}(A,\sigma_{A})$,  and initial phase $0\le\phi_j<A_j$, drawn with uniform probability. The node positions are then perturbed by adding random noise in velocity space as ${\bf n}_j = {\bf n}_{j,0} + {\bf m}$, where each element of the ${\bf m}$ array is randomly drawn from a normal distribution $\mathcal{N}(0,\sigma_{A})$.

From the ensemble of spline models, we evaluated the mean model, $\bar{S}$, and the variance between the models $\sigma^2_{S}$.

\begin{figure}
    \centering
    \includegraphics[width=\linewidth]{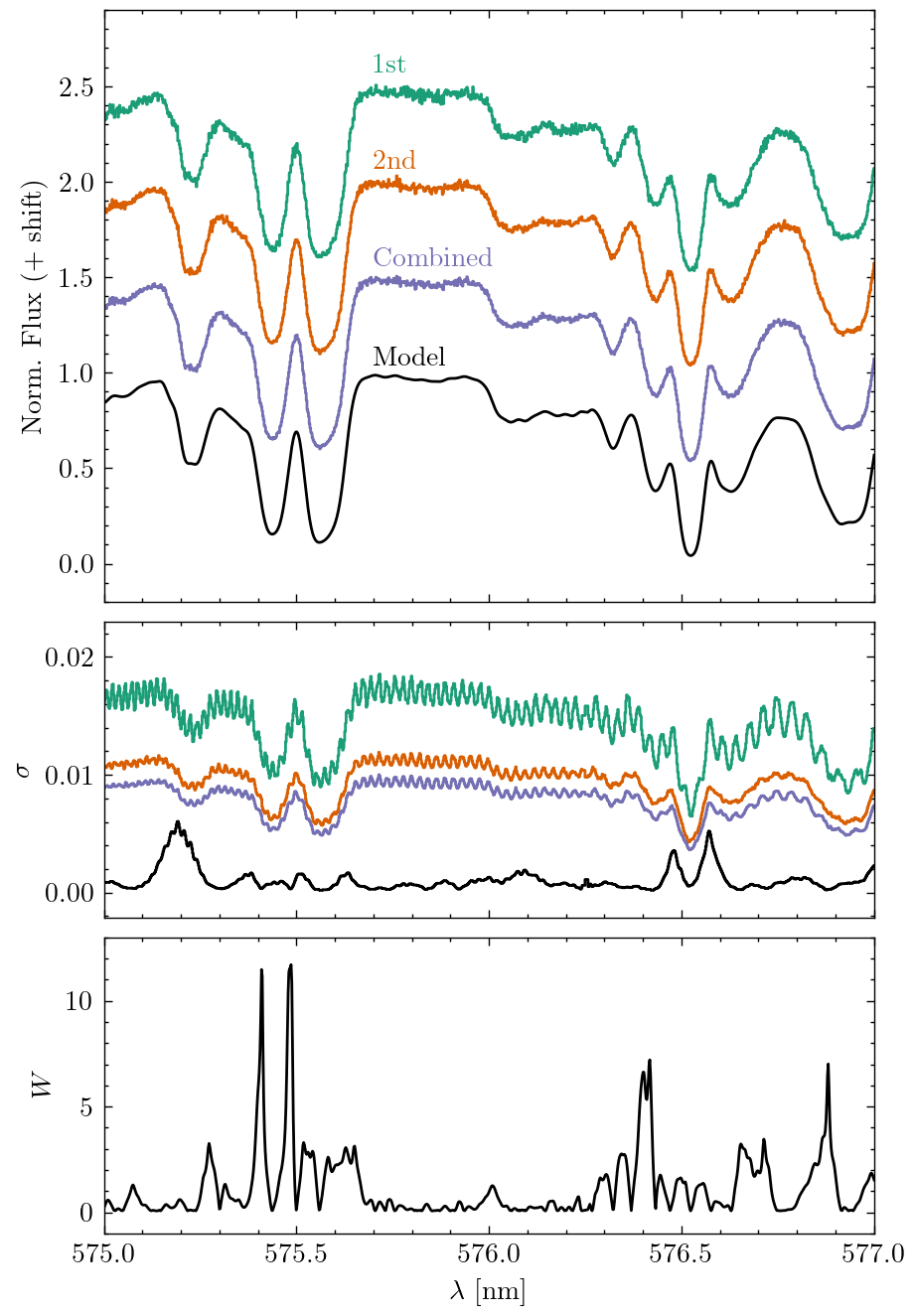}
    \caption{Section of the SB2 spectrum. Top panel: Normalised flux of the 1st epoch (green), 2nd epoch (orange) and combined (purple) spectra. The mean model $\bar{S}$ is shown as a solid black line. The spectra are shifted vertically for clarity. \textit{Middle panel:} The top three coloured lines are the normalised flux errors, $\sigma_F$, for the three spectra shown in the above panel, with the same colour coding. The black solid line is the standard deviation of our model, $\sigma_{\bar{S}}$, i.e.\ it shows where our ensemble of models exhibits large variation and hence uncertainty. Bottom panel: Normalized pixel weights $W$ as defined in Eq. \ref{eq:norm_weights}. Regions with higher values provide more constraints on our final $\dot{z}$ measurement.}
    \label{fig:spec_model}
\end{figure}

Fig.~\ref{fig:spec_model} shows a region of the 1st, 2nd and combined spectra, alongside the mean spline model. In the figure, flux noise and model variance are shown in the middle panel. The periodic pattern found in the flux noise results from the coaddition of several exposures and the subsequent rebinning to a fixed grid, which creates an aliasing effect in the error arrays, whose frequency depends on the position of the pixels within the spectral order. The bottom panel shows the pixel weights used in the measurement (see Eq.~\ref{eq:norm_weights}), proportional to the local model derivative. 

\subsection{Velocity drift estimate}\label{sec:Dv_measure_model}
Strictly speaking, $|\dot{z}|$ is larger at the red end of the Lyman-$\alpha$  forest than at its blue end, causing a redshift dependent compression of features in our spectrum. However, this effect is tiny and completely negligible at the S/N of our data. We therefore ignored the redshift dependence of $\dot{z}$ and assumed that only a rigid translation $\Delta v$, in velocity space, occurred between the two epochs. 
$\Delta v$ was measured indirectly, by first calculating the velocity shifts $\delta v$ occurring between individual epochs and the model. The measured shift between two epochs is then: $\Delta v = \delta v_2 - \delta v_1$, where subscripts identify the epoch. 

 The value of $\delta v$ was obtained through a Monte Carlo Markov Chain (MCMC) process with a modified Gaussian likelihood of the form:
\begin{equation}\label{eq:likelihood}
   \ln \mathcal{L}(\delta v) = 
-0.5 \sum_i^{N_{pix}}{W(\lambda_i|\delta v)
\frac{\left(F_i-\bar{S}(\lambda_i|\delta v)\right)^2}
{\sigma^2_{F,i}+\sigma^2_{\bar{S}}(\lambda_i|\delta v) }},
\end{equation}
where the sum is performed over all non-masked pixels. $F$ and $\sigma_F$ are the normalised flux and its uncertainty, respectively, of either the 1st or 2nd epoch spectra of SB2.
$N_{pix}$ is the number of non-masked pixels, $\bar{S}(\lambda | \delta v)$ is the mean spline model and $\sigma^2_{\bar{S}}(\lambda|\delta v)$ is the variance within the ensemble of spline models, where the latter is included as an additional term at the denominator to effectively consider `less trustworthy' the pixels for which the ensemble's models are not consistent. For both terms, the dependence on $\delta v$ denotes the fact that the splines' nodes have been solidly shifted by $\delta v$ in velocity space, and the models are re-evaluated on the spectrum's pixel grid at each step of the MCMC process. The term $W$ contains information on pixel weights.

Similarly to \citet{Bouchy2001}, we wanted to assign a higher weight to pixels sensitive to velocity shifts, such as those with large flux gradients (i.e.\ the sides of the lines) and those for which variance in the flux gradient is minimal (i.e.\ for which models are consistent). The former aspect is similar to what is done in Sect. \ref{sect:direct}, but the latter aspect was only possible because we produced a large number of spline models. The two contributions were defined by evaluating the mean derivative (in absolute values) $\bar{S}'=\langle\left| dS_j / d\lambda\right|\rangle$ and the standard deviation within first derivatives' magnitudes in the model ensemble $\sigma_{S'}$.
The weights were therefore defined as
\begin{equation}\label{eq:norm_weights}
   W(\lambda_i|\delta v) = N_{pix}\frac{\bar{S}'(\lambda_i|\delta v)}{\sigma_{S'}(\lambda_i|\delta v)}
   \left[\sum_j^{N_{pix}} \frac{\bar{S}'(\lambda_j|\delta v)}{\sigma_{S'}(\lambda_j|\delta v)}\right]^{-1},
\end{equation}
where $\bar{S}'$ and $\sigma_{S'}$ are the mean absolute model derivative and its standard deviation among models. The normalization of the $W$ term is assumed such that $\sum_i^{N_{pix}}W(\lambda_i) = N_{pix}$, as in the non-weighted case ($W(\lambda_i) = 1$ for all pixels). As $\bar{S}$ and $\sigma_{\bar{S}}$, $W$ is also recalculated at each step of the MCMC process. In practice, the new scheme resulted in weighting line edges more strongly than the continuum and other flat regions, as seen in the bottom panel of Fig. \ref{fig:spec_model}.

\begin{figure}
    \centering
    \includegraphics[width=\linewidth]{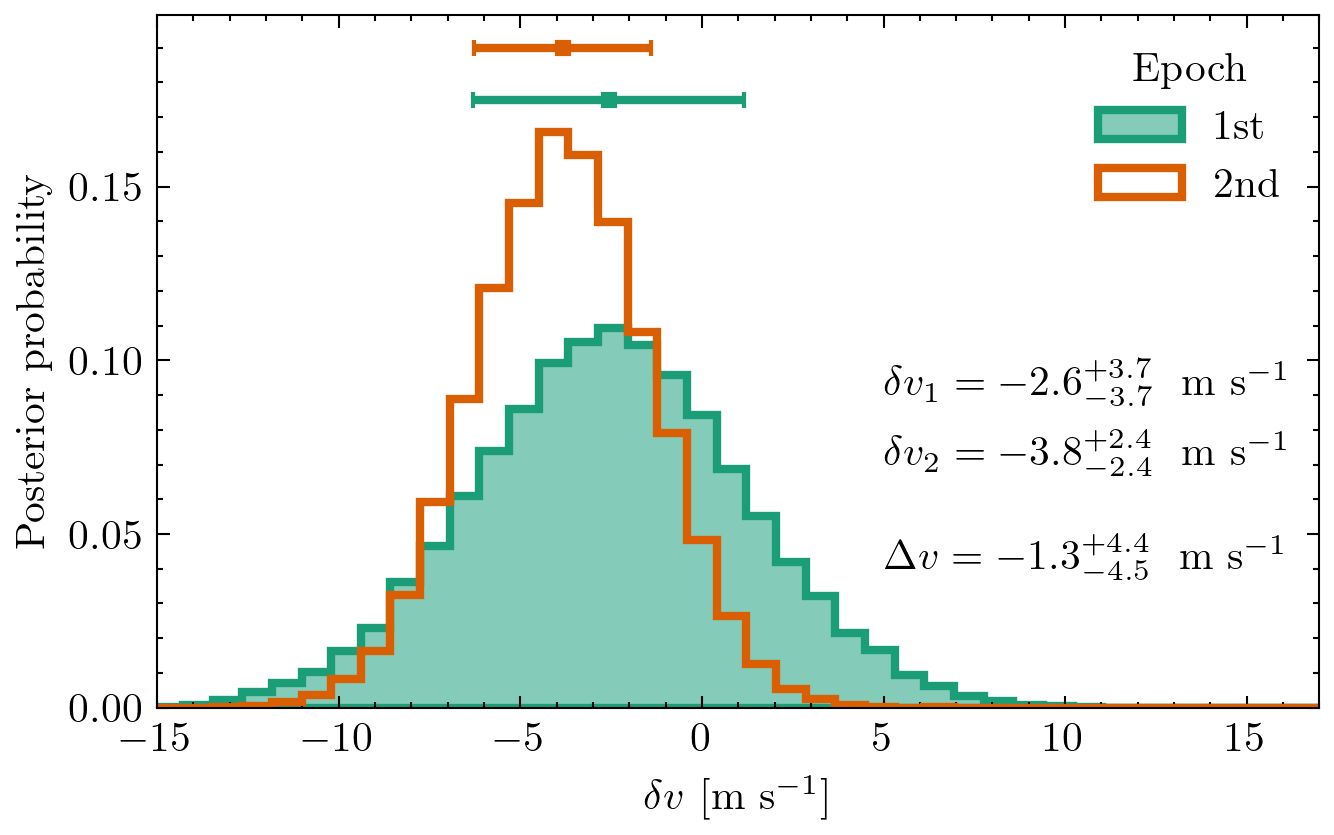}
    \caption{Posterior probability distributions obtained by the MCMC analysis comparing the mean model to the spectra of epoch 1 (green filled) and epoch 2 (orange) through the likelihood defined in Eq.~\ref{eq:likelihood} (see Sect.~\ref{sec:Dv_measure_model}). Scatter points with error bars define the median values and the 68\% confidence intervals. }
    \label{fig:Posteriors_data}
\end{figure}

We ran MCMC to estimate $\delta v$ for each epoch independently. 
Fig.~\ref{fig:Posteriors_data} shows the posterior probability distributions probed by the MCMC analysis for $\delta v_1$ (green filled) and $\delta v_2$ (orange).
The obtained results for the shifts are 
$\delta v_1 = -2.57 _{- 3.73} ^{+ 3.72}  \;{\rm m\,s^{-1}}$ 
and 
$\delta v_2 =-3.82 _{- 2.44}^{ + 2.43} \;{\rm m\,s^{-1}}$, 
where the uncertainties correspond to the 16th and 84th percentiles of the posterior distribution and should be considered equivalent to the 68\%  confidence interval.
Taking the difference between the two values above and propagating their uncertainties, we obtained a velocity shift between the two epochs of 
$\Delta v =  -1.25 _{- 4.46 }^{+ 4.44}\; {\rm m\,s^{-1}}$, which translates into a drift of 
$\dot{v}=-1.43 _{- 5.10 }^{+ 5.08} \; {\rm m\,s^{-1}\,yr^{-1}}$ or 
$\dot{z}=-2.19_{-7.78}^{+ 7.75}\times 10^{-8}\;{\rm yr^{-1}}$.

Both estimates of $\delta v_i$ have negative values. While this may first seem surprising, there is an easily understandable explanation related to the weighting scheme implemented in Eq. \ref{eq:likelihood}. To demonstrate this, we carried out a simple test of measuring $\delta v$ on the combined spectrum, i.e.\ the same spectrum used to produce our 500 models. Na{\"i}vely, this should return a shift consistent with zero. Applying our measurement method to this data resulted in $\delta v_0=-2.71_{ - 2.02}^{ + 2.08} \; {\rm m\,s^{-1}}$, a value that is negative and marginally inconsistent with zero. However, temporarily fixing $W=1$ in Eq. \ref{eq:likelihood} resulted in $\delta v_0= -0.43_{ - 3.51} ^{+ 3.59}$, consistent with zero. Obviously, our choice of $W$ has given more weight to pixels with negative contributions to $\delta v$ while simultaneously decreasing measuring uncertainties. Another way of looking at this is through a comparison of $\delta v_0$ with $\delta v_1$ and $\delta v_2$. Comparing their values, one sees that $\delta v_0$ falls in between $\delta v_1$ and $\delta v_2$, as expected.  
In that sense, the obtained values of $\delta v_1$ and $\delta v_2$ are not surprising.   

\section{Validation and systematics}\label{sec:validation}
\subsection{Measurement validation}
We validated the measurement procedures by applying the pixel-by-pixel (see Sect.~\ref{sect:direct}) and the model-based (see Sect.~\ref{sec:Dv_measure_model}) methods to a synthetic sample of mock spectra for which we know a priori the velocity shift occurring between the two epochs. We thus produced mock datasets assuming the same S/N distribution, pixel size, spectral range and resolution of the SB2 spectra, simulating a $\Delta v$ of $0~\cms$, $-0.38~\cms$, $-4.3~\ms$, and $-42.9~\ms$, corresponding to a temporal baseline of zero, $0.875$, $10^3$ and $10^4$ years in a $\Lambda$CDM~Universe.

\begin{figure*}
    \centering
    \includegraphics[width=\linewidth]{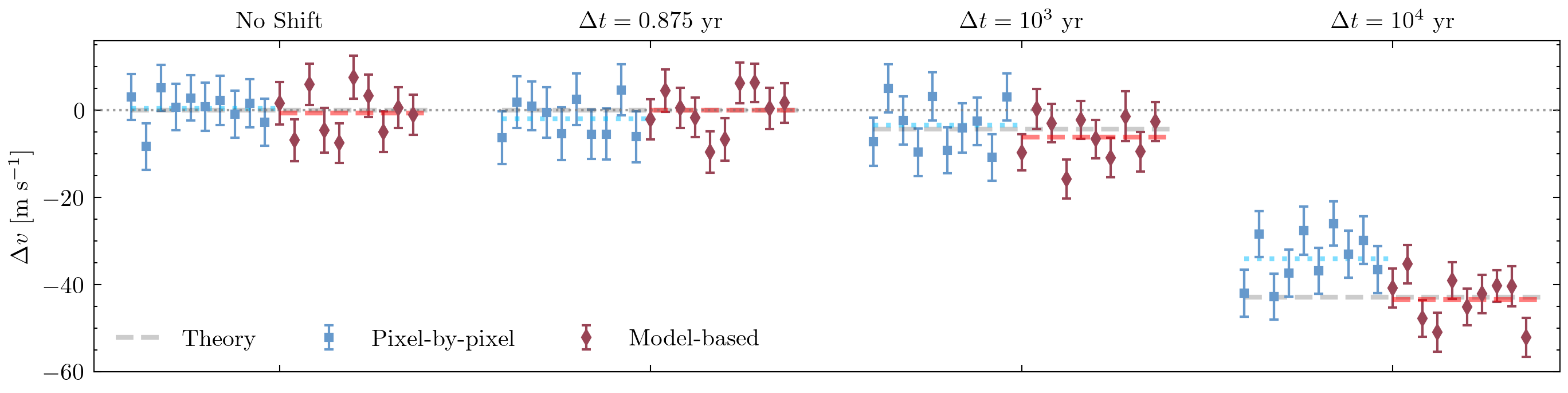}
    \caption{Velocity shift $\Delta v$ measured on 10 random mock sightline pairs with both the pixel-by-pixel (blue square scatter points, see Sect.~\ref{sect:direct}) and model-based methods (red diamond scatter points, see Sect.~\ref{sec:Dv_measure_model}) where a different baseline is assumed between the two epochs. The horizontal blue dotted and red dashed lines report the average $\Delta v$ measured with the pixel-by-pixel and the model-based methods, respectively  From left to right, the mock datasets are built assuming no shift ($\Delta v_{\rm exp} = 0~\ms$), $\Delta t=0.875 ~{\rm yr}$ (the same time elapsed between the two observed spectra, $\Delta v_{\rm exp} = -0.38 ~\cms$), $\Delta t=10^3 ~{\rm yr}$ ($\Delta v_{\rm exp} = -4.3 ~\ms$), and $\Delta t=10^4 ~{\rm yr}$ ($\Delta v_{\rm exp} = -42.9 ~\ms$). The grey dashed lines report the assumed shift.}
    \label{fig:mocks_measure}
\end{figure*}

We applied our methods to 10 random sightlines from each mock dataset, with the results shown in Fig.~\ref{fig:mocks_measure}. 
The top panel shows the measured velocity shifts $\Delta v$ obtained with the pixel-by-pixel (blue) and model-based (red) methods in the four baseline cases. Both methods were applied to the same set of 10 sightlines.
We found that the two methods are capable of recovering the imposed drift, with an uncertainty that depends solely on the S/N level of the two spectra (matched to the SB2 data, regardless of the simulated $\Delta v$) and on the number of Lyman-$\alpha$  lines contained in each sightline. The scatter among the 10 measured velocity shifts is consistent with the estimated uncertainties. 
Notably, the pixel-by-pixel method shows a slight discrepancy between the expected signal and the mean measured $\Delta v$ when applied to the longest baseline of 10,000 years. 
The cause of such bias is not clear, although it could stem from the perturbative assumption taken in Eq.~\ref{eq:gradiente_method} that might not hold any longer for the large shift imposed in this simulated case. Nonetheless, our aim is to measure very small shifts for which such an assumption is solid.
The model-based method recovers the imposed shift well in all instances. 

\subsection{Uncertainty across the spectrum}

It is possible that a narrow segment of the Lyman-$\alpha$  forest contains unidentified contaminants that spoil the measurement on $\Delta v$. We investigated this possibility by dividing the Lyman-$\alpha$  forest into 10 equally sized chunks (of width $9.4~{\rm nm}$) and measuring $\Delta v$ over each individual one, using both the pixel-by-pixel and model-based methods, following the usual procedures described in Sects.~\ref{sect:direct} and \ref{sec:Dv_measure_model}.

\begin{figure}
    \centering
    \includegraphics[width=\linewidth]{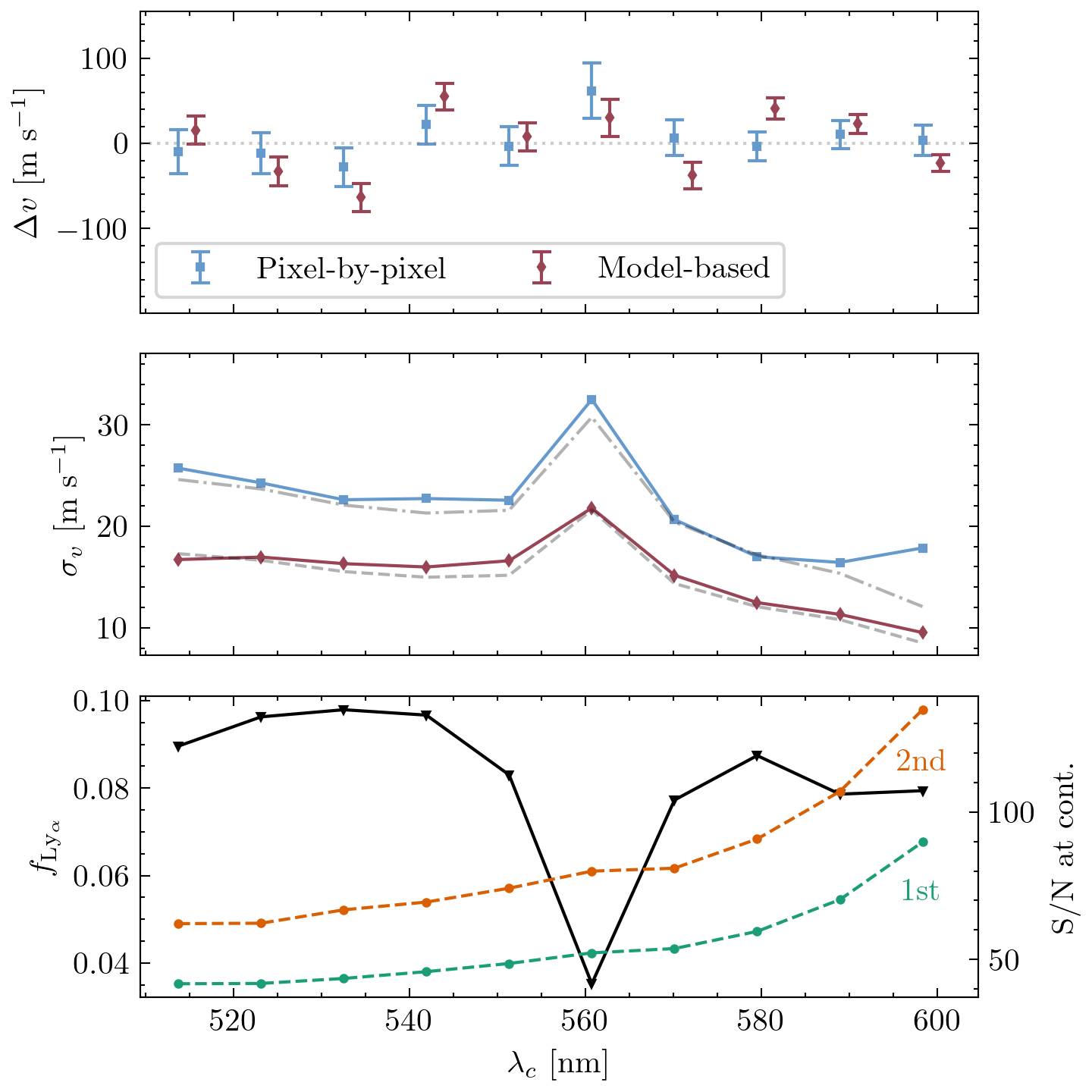}
    \caption{Estimated velocity shift $\Delta v$ computed on 10 equispaced sections of the Lyman-$\alpha$  forest. Top panel: $\Delta v$ estimated on each spectral chunk with the pixel-by-pixel (blue squares) and model-based (red diamonds) methods, as a function of the section's central wavelength $\lambda_c$. Middle panel: measurement uncertainty $\sigma_v$ for the two methods, shown with the same colour coding as the panel above. The dot-dashed and dashed grey lines report the expected uncertainty of the two methods, based on the median S/N and pixel number of each chunk. Bottom panel: the solid black line and the left-hand axis denote the fraction of the Lyman-$\alpha$  forest $f_{\rm Ly_{\alpha}}$ considered in the measurement within each chunk, taking into account the masking of metals, strong absorbers and bad pixels. Note that the spectral chunk centred at $560.7~{\rm nm}$ contains the sub-DLA absorber and has a significantly smaller fraction of usable pixels than the other regions. The right-hand axis of the same panel reports the median S/N at continuum level in each chunk in the 1st and 2nd epoch spectra, shown with dashed green and orange lines, respectively.}
    \label{fig:chunky}
\end{figure}

Fig.~\ref{fig:chunky} shows the measured $\Delta v$ for each chunk, as a function of the chunk's central wavelength $\lambda_c$, in the upper panel.
The middle panel shows the velocity shift uncertainty $\sigma_v$ computed on each chunk with the pixel-by-pixel (blue) and model-based (red) methods. 
In the absence of systematics, the measurement uncertainty is expected to depend solely on the median S/N at continuum and the number of pixels (or the fraction of the forest $f_{\rm Ly_{\alpha}}$) in each chunk, scaling as $\sigma_v\propto{\rm S/N}^{-1}  f_{\rm Ly_{\alpha}}^{-0.5}$. 
The expected measurement uncertainty is shown in the middle panel as a dot-dashed (dashed) grey line, scaled by the pixel-by-pixel (model-based) uncertainty obtained on the full spectrum. The estimated uncertainties closely follow the scaling expectations for both methods, highlighting the absence of wavelength-dependent systematic effects.

\subsection{Influence of the quasar}\label{sec:quasar_variability}
Given the extraordinary nature of SB2 \citep{Wolf2024}, with its high black hole accretion rate ($\dot{M}\sim370 ~M_{\odot} {\rm yr^{-1}}$) and luminosity variability ($\sim 15\%$ over the last 6 years), one would expect to find a strong signature of IGM photoionization in the Lyman-$\alpha$  forest close to the quasar emission redshift. However, our data do not show an important decrease in strength and number of Lyman-$\alpha$  lines close to $z_{em}$, nor a significant variability of such lines, which could be related to the quasar's strong activity. Therefore, in our previous analysis, we based the redshift drift measurement on the whole forest, without excluding the proximity region, about $5000~\kms$ from the quasar emission redshift. 

We checked for systematic effects due to the proximity to the quasar by performing the same measurement as before, but excluding such region, and measured a velocity drift between the two epochs of $\Delta v = 3.88\pm 5.08~\ms$. Having excluded the proximity region reduces the number of pixels used in the measurement to $\sim 37400$, or $\sim 74\%$ of the total amount found in the forest. When ignoring the proximity region, we excluded the range close to the Lyman-$\alpha$  emission where the spectra have the highest S/N levels at the continuum, with a median of $89$, $134$ and $161$ per pixel in the 1st epoch, 2nd epoch and combined spectra, respectively, and performed the measurement on spectra with effectively lower S/N. Therefore, the uncertainty grows not only due to a reduced pixel sample, but also due to a smaller median S/N level ($\sim 82$ per pixel at the continuum in the combined spectrum) outside of the proximity region. These two effects add up to a scaling factor $\left(82/86\right)^{-1}  \left( 0.74/0.84\right)^{-1/2}\sim1.12$, explaining the new uncertainty.

Still, the measurement is compatible with the expected value, and no systematic effect due to excluding or taking into account the proximity region is clearly visible.

\subsection{Influence of local motions}\label{sec:local_motions}

As shown by \cite{Inoue2020}, the mass of the Milky Way (MW) and the other galaxies in the Local Group (LG), especially the Large Magellanic Cloud (LMC) and M31, have a non-negligible impact on the proper motion of the Sun, providing a local source of acceleration along a certain sightline. Left uncorrected, this effect will impact the cosmological redshift drift signal measurement. The amplitude of this effect is sightline dependent and, in the case of SB2, induces an additional systematic drift in the Lyman-$\alpha$  lines of $\dot{v}_{\rm \odot}=-0.16\pm0.04\;\cmsyr$. This is of the same order of magnitude as the cosmological signal we want to measure.
Thus, the expected velocity shift between two epochs of SB2 grows effectively to $\dot{v}_{\rm SB2}\sim-0.59~\cmsyr$, implying it can be measured on a shorter temporal baseline, at fixed total integration time. 

However, such shift is the composite of the local and cosmological effects, where the former should be removed from the measured shift to infer the actual redshift drift. 
The required duration of the experiment is not affected by the local component, if not for the fact that its amplitude is inferred from measurements of the MW and LG's galaxies masses, and has an uncertainty that will be propagated to the redshift drift measurement's uncertainty. Such uncertainty is one order of magnitude smaller than the expected redshift drift, of order $0.04~\cmsyr$, where the cosmological signal is $\sim -0.429~\cmsyr$, and its propagation in the actual measurement is sub-dominant with respect to the measurement uncertainty related to the spectral analysis (see Sect.~\ref{sec:future}).

\subsection{Instrumental systematics}\label{sec:instrumental_systematics}
ESPRESSO's wavelength calibration is a crucial aspect for robustly measuring the redshift drift signal. Previous studies identified differences as large as $20~\ms$ between the two calibration methods available on ESPRESSO, the LFC and the Fabry-P{\'e}rot (FP) etalon combined with a ThAr lamp \citep{Schmidt2021}. Their impact is expected to decrease by considering a large number of lines over a broad spectral range, but the true quantification on a real measurement was so far lacking. We therefore repeated our analysis on SB2 observations calibrated using ThAr+FP frames instead of LFC, created specifically for this purpose following the same procedures used previously in Sect.~\ref{sec:model_build}-\ref{sec:Dv_measure_model}. We measured $\Delta v = -0.81_{-4.39}^{+4.47}~\ms$ from these data, a change of $+0.44~\ms$ with respect to LFC-calibrated data. The two measurements agree within uncertainties. 

Another source of uncertainty relates to assumptions on the shape of the line-spread function (LSF), used for determining line centres during wavelength calibration. The standard ESPRESSO DRS wavelength calibration procedures assume a Gaussian LSF shape and, because this cannot be modified without significant changes to the entire procedure, we were unable to directly quantify the impact of this assumption on our results. However, considering that using a Gaussian LSF should introduce wavelength calibration errors similar to the differences observed between using LFC and ThAr+FP (i.e. $20~\ms$, \cite{Schmidt2021, Schmidt2024}), a reasonable expected additional systematic uncertainty is $\sim 0.5~\ms$ through analogy with the previous paragraph. 

\section{Future perspectives}\label{sec:future}

Based on our results, we estimated the time required for a statistically significant detection of the cosmological signal with ESPRESSO and ANDES.

We envision an observational monitoring programme, carried out for $N_e$ epochs, with a cadence of $T$ hours of ESPRESSO integration of SB2 each year.
The precision on the measurement of the cosmic acceleration ($\dot{v}\sim\Delta v / \Delta t$) scales in time, adapting Eq.~\ref{eq:vel_uncertainty_liske} to account for the additional temporal dependency, as
\begin{equation}\label{espressotime}
    \sigma_{\dot{v}} \propto g(N_e)\times{\rm S/N_{tot}}(N_e, T)^{-1}\times\left( N_e - 1\right)^{-1},
\end{equation}
where the experiment baseline is $\Delta t = N_e - 1$ (e.g. 1 year of programme has 2 epochs, 2 years of programme have 3 epochs, and so on), and the total S/N obtained after $N_e$ epochs scales as ${\rm S/N_{tot}}(N_e, T)\propto\sqrt{N_eT}$. The form factor $g(N_e)$ depends on the distribution across time of the observational epochs, being 1.7 for the continuous monitoring program we considered. 

\begin{figure}
    \centering
    \includegraphics[width=\linewidth]{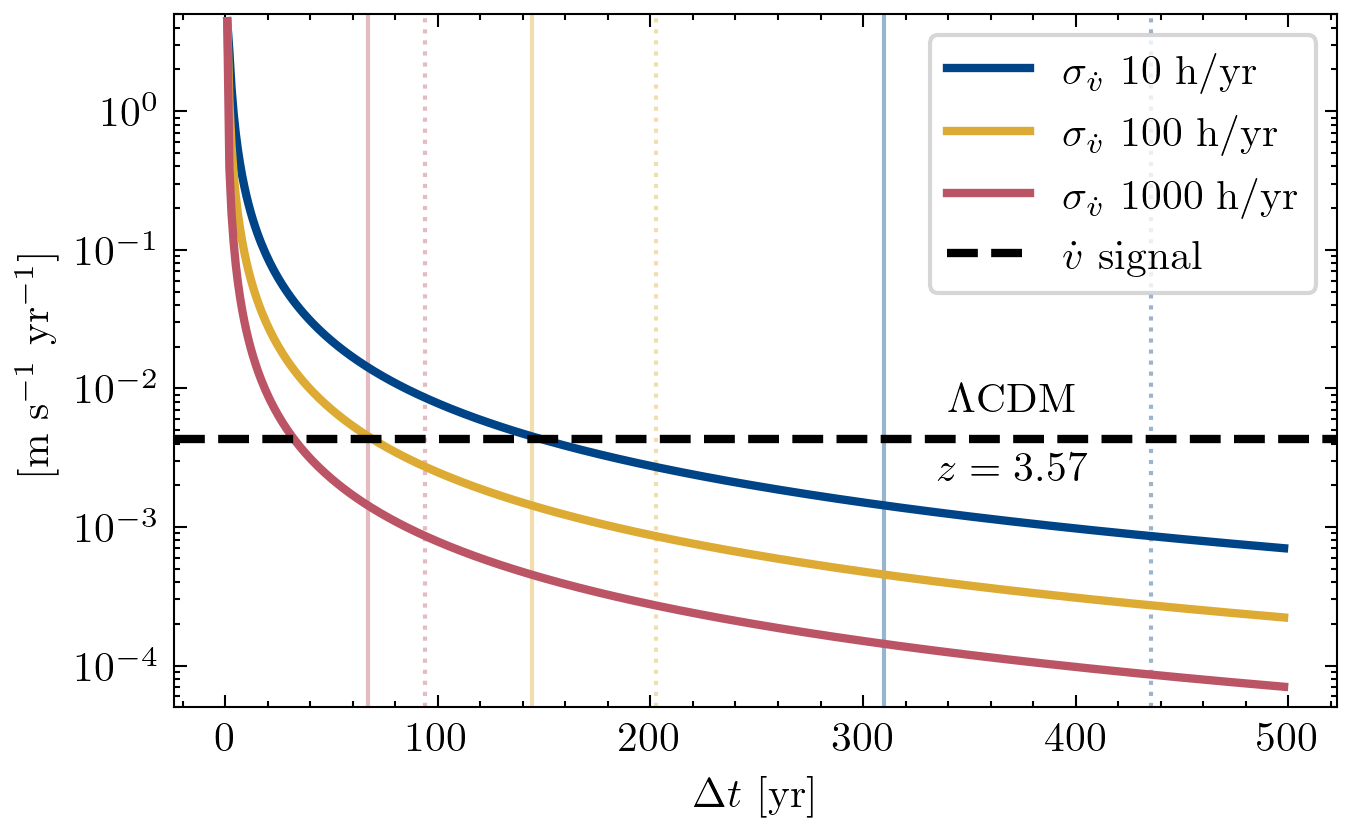}
    \caption{Velocity shift uncertainty reached with ESPRESSO spectra of SB2, as a function of total temporal baseline of the experiment, assuming three different observational strategies with an integration time of 10 hours per year (blue), 100 hours per year (yellow), and 1000 hours per year (red). The horizontal dashed black line defines the magnitude of the cosmological signal expected for a $\Lambda$CDM~ Universe at the average redshift of SB2's Lyman-$\alpha$  forest. Vertical lines define the time needed to achieve a 3$\sigma$ (solid) and 5$\sigma$ (dotted) detection for the three observational strategies.}
    \label{fig:unce_future}
\end{figure}

Fig.~\ref{fig:unce_future} shows the acceleration measurement uncertainty $\sigma_{\dot{v}}$ achieved with ESPRESSO as a function of the temporal baseline of our experiment, starting from the data already collected, and following up with a monitoring campaign of 10, 100, or 1000 hours of integration time each year. The vertical solid lines denote at what epoch we can achieve a 3$\sigma$ (or 5$\sigma$ vertical dotted lines) detection of the cosmological signal in the three cases. The estimates of the required baseline for a detection are also reported in Table~\ref{tab:unce_future}.

\begin{table}[]
    \centering
    \caption{Temporal baselines required to achieve a 1$\sigma$, 3$\sigma$ or 5$\sigma$ detection following up the SB2 spectra with a total integration time of 10, 100, or 1000 hours per year, if performed with ESPRESSO or ANDES (assuming a $10\%$ efficiency).}
    \begin{tabular}{lrrr}\hline
           T [h/yr] & 1$\sigma$ & 3$\sigma$ & 5$\sigma$ \\ \hline
           \multicolumn{4}{c}{ESPRESSO} \\ \hline
         10  & 149 yr & 310 yr & 436 yr\\
         100  & 70 yr & 145 yr & 203 yr\\
         1000  & 33 yr & 68 yr & 95 yr \\ \hline
         \multicolumn{4}{c}{ANDES} \\ \hline
         10  & 56 yr & 115 yr & 161 yr\\
         100  & 26 yr & 54 yr & 75 yr\\
         1000  & 13 yr & 25 yr & 35 yr \\ \hline
    \end{tabular}

    \label{tab:unce_future}
\end{table}

Clearly, the timescales for the experiment are long due to the limited collecting area of VLT. However, these results can be used to estimate, at first order, what will be the impact of moving the observational campaign to the ELT/ANDES spectrograph \citep{Marconi2024}. Assuming that the total efficiencies of the ESPRESSO and ANDES spectrographs are similar (about $10\%$), since the latter is fed by a telescope with a collecting area $\sim 20$ times larger than the area of one VLT unit, the same accuracy in measuring $\Delta v$ can be reached in $1/20$-th of the total ESPRESSO integration time. 
By this logic, a 3$\sigma$ (5$\sigma$) detection of the redshift drift can be carried out observing SB2 with ANDES in 68 years (95 years), with 50 hours of total integration time per year, instead of 1000 hours per year as in the case of ESPRESSO.

Similarly, assuming a fixed yearly integration time, the same level of precision in the measurement of the cosmic acceleration $\sigma_{\dot{v}}$ can be achieved with ANDES with a temporal baseline that is a factor $20^{1/3}\sim2.714$ shorter than in the case of an ESPRESSO experiment, due to the scaling of Eq.~\ref{espressotime}.
In this case, a 3$\sigma$ (5$\sigma$) detection of the redshift drift can be carried out by means of a monitoring campaign of SB2 with 100 hours of integration per year, in 54 years (75 years) with ANDES, instead of 145 years (203 years) with ESPRESSO.
Table~\ref{tab:unce_future} reports the temporal baselines required for detection in an ANDES-based experiment, when allocating 10, 100, or 1000 total hours of exposure per year. 

The ANDES spectrograph is yet to be constructed, and its efficiency could change in the future design phases. We can easily parametrise the required baseline of the measurement to reach a 3$\sigma$ detection, letting the total efficiency of ANDES, $\epsilon$, as a free parameter:
\begin{equation}\label{eq:delta_t_ANDES}
    \Delta t^{\rm ANDES}_{3\sigma} \approx 54\; \left(\frac{T}{100~{\rm h\; yr^{-1}}}\right)^{-1/3} \left(\frac{\epsilon}{0.1} \right)^{-1/3}~{\rm yr},
\end{equation}
where $T$ is the total integration time of an ANDES-like spectrograph per year of the experiment. The total amount of integration hours needed for a 3$\sigma$ detection is simply
\begin{equation}\label{eq:total_integration_time}
    {\rm Total~integration~time} = 5.4\times 10^3 \left(\frac{T}{100~{\rm h\; yr^{-1}}}\right)^{2/3} \left(\frac{\epsilon}{0.1} \right)^{-1/3}~{\rm h},
\end{equation}
depending on the amount of telescope time that can be continuously allocated to this project.

\section{Results and discussion}\label{sec:discussion}

We set out to assess the feasibility of measuring the redshift drift signal using Lyman-$\alpha$ forest spectroscopy, based on the highest-quality data available prior to the commissioning of ELT/ANDES. We used new, high-resolution and high-S/N spectra of the most luminous known quasar in the Universe, J052915.80-435152.0 (SB2; $z_{\rm em}=3.962$), obtained with ESPRESSO at the VLT. This exceptionally bright object is especially suited to perform the Sandage test of cosmic redshift drift, owing to its luminosity and emission redshift. We began this decades-long experiment with the ultra-stable spectrograph ESPRESSO over a one-year baseline. The primary goal was to investigate the measurement's intrinsic systematics and validate the predictions on the expected total integration time required to achieve a statistically significant detection. This work presents the first two epochs of the experiment, intended to serve as an anchor for comparison to future observations, initially carried out with ESPRESSO and subsequently with ELT/ANDES. 

Our main results are:
\begin{itemize}

    \item Applying the pixel-by-pixel method of \citet{Bouchy2001} (Sect.\ \ref{sect:direct}), we measured a velocity shift between two epochs of $\Delta v = -2.67 \pm 6.64~\ms$. These uncertainties are 65\% above the prediction made by \citet{Liske08} ($\sigma_{v, {\rm Liske}} = 4.02~\ms$, see Eq.~\ref{eq:vel_uncertainty_liske}). 

    \item Using an ensemble of 500 models specifically tailored to the resolution, pixel size and S/N of our data (Sect.\ \ref{sec:model_based}), we identified spectral regions more sensitive to velocity shifts and gave them more weight in our analysis, obtaining $\Delta v = -1.25_{- 4.46}^{+ 4.44}\; {\rm m\,s^{-1}}$. This is a 30\% improvement in statistical uncertainty over the direct method, and only $10\%$ greater than the value predicted by \citet{Liske08}.
     
    \item The redshift drift measured between the two epochs turns out to be $\dot{z}= -2.19 _{- 7.78} ^{+ 7.75}\times 10^{-8}\;{\rm yr^{-1}}$ at $\langle z \rangle =3.57$, which represents the main result of this paper. This value is fully consistent with the value expected within the $\Lambda$CDM~ model, $-6.6\times 10^{-11}\;{\rm yr^{-1}}$. The quoted uncertainties are only statistical and three orders of magnitude larger than the cosmological signal. Systematic uncertainties due to the motion of the Solar system, temporal variability of SB2's luminosity, and wavelength calibration are currently subdominant. 
    
    \item Based on these findings, we estimate that a $3\sigma$ detection of the redshift drift could be achieved with ESPRESSO (ANDES) after 145 (54) years, assuming 100 hours of observation per year for SB2 (Table\ \ref{tab:unce_future}). 
   
\end{itemize}
The best current upper limit on cosmic acceleration is $|\dot{v}|\leq2.2\; {\rm m\,s^{-1}\,yr^{-1}}$ at $1\sigma$ confidence level \citep{Darling2012}. Our measurement of $\Delta v = -1.25_{- 4.46}^{+ 4.44}\; {\rm m\,s^{-1}}$ translates into $\dot{v}=-1.43_{- 5.10} ^{+ 5.08} \; {\rm m\,s^{-1}\,yr^{-1}}$ (Sect.\ \ref{sec:model_based}), making our limit a factor $2.3$ larger than that of \citet{Darling2012}. 
However, their result is based on observations of H~\textsc{i} 21 cm absorption line systems in 10 objects over a baseline of 13 years, while we obtained a comparable result with only 12.4h of Lyman-$\alpha$  forest and a $\sim1$ year baseline. By expanding our dataset by another 12 hours over the next 1.5 years, we expect to match the uncertainty in the measurement of \citet{Darling2012}, showcasing the power of Lyman-$\alpha$  observations for $\dot{z}$ measurements. 

Interestingly, neither the pixel-by-pixel method (Sect.\ \ref{sect:direct}) nor our newly developed method based on splines (Sect.\ \ref{sec:model_based}) precisely reproduce the uncertainties predicted by the scaling relation of \citet{Liske08}. This discrepancy may arise because the \citet{Liske08} relation was calibrated assuming a measurement carried out on a sample of multiple objects, whereas we analysed a single sightline.  Sightline-to-sightline fluctuations in the measurement uncertainties are expected due to cosmic variance, where the analysis of SB1 will provide further insight into this issue (Marques et al., in prep.).
    
We examined the validity of our methods and some potential sources of systematic uncertainties in Sect.\ \ref{sec:validation}. While both the direct and the spline methods recover simulated $\Delta v$ (within uncertainties) for $|\Delta v|<5~\ms$, the pixel-by-pixel methods shows a systematic underestimation of the shift when recovering the simulated $|\Delta v|\approx40~\ms$ (Fig.\ \ref{fig:mocks_measure}). Such a result is not currently understood. Solar system accelerations add $\sim 1~\cmsyr$ to the expected signal, being negligible at the current level of S/N.
The presence of such effect does not alter the experiment's expected temporal baseline (see Sect.~\ref{sec:local_motions}). 
Excluding the quasar proximity zone from our analysis shifted $\Delta v$ by $5.14~\ms$, a value only marginally inconsistent ($\sim1.01\sigma$) with the statistical uncertainty (Sect.~\ref{sec:quasar_variability}). The shift is likely a consequence of a large (10\%) decrease in the number of available pixels, and not a true estimate of uncertainty from quasar variability. More data and longer time baselines are needed to confirm whether temporal variability in quasar flux causes a systematic effect on $\Delta v$. 
Wavelength calibration uncertainties are estimated to be $\approx1~\ms$ (Sect.~\ref{sec:instrumental_systematics}). Considering the latter is approximately one fourth of the statistical uncertainty, we expect that systematic effects will start playing a significant role in the error budget (in SB2 measurements) once the total S/N quadruples (c.f.\ Eq.~\ref{eq:vel_uncertainty_liske}). 

A caveat of both methods used in this paper is that they are unable to disentangle a real velocity shift in the data from any changes to the instrument properties, such as its LSF. LSF is known to change when important components are replaced \citep{LoCurto2015Msngr.162....9L}, and is likely to occur in experiments spanning several decades. There were no major interventions in ESPRESSO over the time period spanning observations presented here that would raise concerns about LSF or other instrument properties varying. However, removing instrumental imprints will be important in the future, for example by using advanced data reduction frameworks such as `spectro-perfectionism' \citep{Bolton2010PASP..122..248B} or similar \citep{Piskunov2021A&A...646A..32P}. 

Based on considerations presented in Sect.\ \ref{sec:future}, a $3\sigma$ detection of $\dot{z}$ using SB2 will require 54 years of observations with ANDES when allocating 12.5 nights per year to this project (Table\ \ref{tab:unce_future}). Shorter timescales can be achieved with a more intensive campaign or by increasing the total efficiency of ANDES (c.f.\ Eq.~\ref{eq:delta_t_ANDES}). For obvious reasons, an intensive monitoring campaign of such magnitude focusing on only one object is not ideal. In a more realistic case, multiple bright objects will be observed simultaneously, spread over a large range of right ascensions. A Golden Sample of 7 quasars compiled with this goal in mind has already been proposed by \citep{Qubrics23}.
In such a case, being SB2 the brightest quasar of the sample, the experiment baseline will slightly increase with respect to Eq.~\ref{eq:delta_t_ANDES} at fixed total observational cadence $T$, with time shared among multiple targets.
However, the other quasar's sightline might be better suited for the test, with more narrow lines, fewer strong absorbers and a larger fraction of the forest available for the measurement, than what we found on SB2. A measurement on such sightlines would have smaller uncertainties and counterbalance the lower S/N related to the fainter objects. 
At this level, it is not straightforward to formally predict the consequences of such effects on the baseline of an experiment involving all seven quasars of the Golden Sample.  In a parallel paper (Marques et al, in prep.), we will analyse the spectrum of SB1, the second brightest object of the sample, in order to address this point.

\begin{acknowledgements}
AT is grateful to Vid Ir\v si\v c for the prolific discussions.
The INAF authors acknowledge financial support of the Italian Ministry of Education, University, and Research
with PRIN 201278X4FL and the "Progetti Premiali" funding scheme.
This work was financed by Portuguese funds through FCT (Funda\c c\~ao para a Ci\^encia e a Tecnologia) in the framework of the project 2022.04048.PTDC (Phi in the Sky, DOI 10.54499/2022.04048.PTDC). CJM also acknowledges FCT and POCH/FSE (EC) support through Investigador FCT Contract 2021.01214.CEECIND/CP1658/CT0001 (DOI 10.54499/2021.01214.CEECIND/CP1658/CT0001). CMJM is supported by an FCT fellowship, grant number 2023.03984.BD.
MTM acknowledges the support of the Australian Research Council through Future Fellowship grant FT180100194 and through the Australian Research Council Centre of Excellence in Optical Microcombs for Breakthrough Science (project number CE230100006) funded by the Australian Government.
TMS acknowledges the support from the SNF synergia grant CRSII5-193689 (BLUVES). 
The work of KB is supported by NOIRLab, which is managed by the Association of Universities for Research in Astronomy (AURA) under a cooperative agreement with the U.S. National Science Foundation.
EP acknowledges financial support from the Agencia Estatal de Investigaci\'on of the Ministerio de Ciencia e Innovaci\'on MCIN/AEI/10.13039/501100011033 and the ERDF “A way of making Europe” through project PID2021-125627OB-C32, and from the Centre of Excellence “Severo Ochoa” award to the Instituto de Astrofisica de Canarias.
ASM and JIGH acknowledge financial support from the Spanish Ministry of Science, Innovation and Universities (MICIU) projects PID2020-117493GB-I00 and PID2023-149982NB-I00.
NS and NN acknowledge financial support by FCT - Fundação para a Ciência e a Tecnologia through national funds by these grants: UIDB/04434/2020 DOI: 10.54499/UIDB/04434/2020, UIDP/04434/2020 DOI: 10.54499/UIDP/04434/2020.

\end{acknowledgements}


\bibliographystyle{aa}
\bibliography{AT}

\appendix{}
\onecolumn

\section{Metal lines}
\begin{longtable}{ll ll ll ll l}
	\caption{Parameters obtained from the fit of all detected metal transitions with $\lambda>603~{\rm nm}$.
    The first column reports the ionic transitions. The second and third columns report the redshift position of the Voigt profile with its uncertainty. Fourth and fifth columns contain the logarithmic column density and its uncertainty. Sixth and seventh columns report the Doppler b-parameter and its uncertainty.
    }\\
    \label{tab:metal_list}\\
	\toprule
  \multicolumn{1}{c}{Transition} &  \multicolumn{1}{c}{$z$} & \multicolumn{1}{c}{$\sigma_z$} & \multicolumn{1}{c}{$\log\left(N/{\rm cm^{-2}}\right)$} & \multicolumn{1}{c}{$\sigma_{\text{logN}}$} & \multicolumn{1}{c}{b} & \multicolumn{1}{c}{$\sigma_{\text{b}}$} \\
 &  &  \multicolumn{1}{c}{$[1\times10^{-5}]$} &&& \multicolumn{1}{c}{$[\unit{\kms}]$} & \multicolumn{1}{c}{$[\unit{\kms}]$} \\
        \midrule
	\endfirsthead
	\caption{Parameters obtained from the fit of all detected metal transitions with $\lambda>603~\unit{\nano\meter}$.
    The first column reports the ionic transitions. The second and third columns report the redshift position of the Voigt component with its uncertainty. Fourth and fifth columns contain the logarithmic column density and its uncertainty. Sixth and seventh columns report the Doppler b-parameter and its uncertainty.
    }\\	\toprule
 \multicolumn{1}{c}{Transition} &  \multicolumn{1}{c}{$z$} & \multicolumn{1}{c}{$\sigma_z$} & \multicolumn{1}{c}{$\log\left(N/{\rm cm^{-2}}\right)$} & \multicolumn{1}{c}{$\sigma_{\text{logN}}$} & \multicolumn{1}{c}{b} & \multicolumn{1}{c}{$\sigma_{\text{b}}$} \\
 &   &\multicolumn{1}{c}{$[1\times10^{-5}]$} & && \multicolumn{1}{c}{$[\unit{\kms}]$} & \multicolumn{1}{c}{$[\unit{\kms}]$} \\
    \midrule
	\endhead
	\midrule 
	\multicolumn{9}{r}{\footnotesize\itshape Continue on the next page}
	\endfoot
	\bottomrule
	\endlastfoot
Mg \textsc{i} 2852 & 1.133524 & 0.46 & 10.515 & 0.106 & 5.648 & 1.697 \\
Mg \textsc{i} 2852 & 1.133604 & 0.10 & 11.150 & 0.018 & 4.640 & 0.192 \\
Mg \textsc{ii} 2796, 2803 & 1.444199 & 0.13 & 12.100 & 0.010 & 9.764 & 0.275 \\
Fe \textsc{ii} 2600 & 1.444199 & 0.13 & 11.459 & 0.094 & 7.855 & 2.299 \\
Mg \textsc{i} 2852 & 1.444208 & 0.22 & 10.350 & 0.071 & 2.537 & 0.480 \\
Mg \textsc{i} 2852 & 1.444252 & 0.75 & 11.093 & 0.027 & 18.394 & 1.416 \\
Mg \textsc{ii} 2796, 2803 & 1.444316 & 0.20 & 11.333 & 0.053 & 4.221 & 0.502 \\
Mg \textsc{i} 2852 & 1.444577 & 0.31 & 10.486 & 0.044 & 4.454 & 0.571 \\
Mg \textsc{ii} 2796, 2803 & 1.444590 & 0.47 & 11.667 & 0.111 & 7.081 & 1.214 \\
Mg \textsc{ii} 2796, 2803 & 1.444682 & 0.43 & 11.420 & 0.131 & 4.778 & 0.977 \\
Mg \textsc{ii} 2796, 2803 & 1.444834 & 0.16 & 12.536 & 0.064 & 4.065 & 0.138 \\
Fe \textsc{ii} 2382, 2600 & 1.444834 & 0.16 & 11.794 & 0.042 & 3.151 & 0.477 \\
Mg \textsc{i} 2852 & 1.444838 & 0.19 & 10.908 & 0.022 & 5.629 & 0.348 \\
Mg \textsc{ii} 2796, 2803 & 1.444880 & 0.13 & 11.715 & 0.017 & 4.997 & 0.218 \\
Mg \textsc{ii} 2796, 2803 & 1.444911 & 0.68 & 12.183 & 0.067 & 17.595 & 2.163 \\
Mg \textsc{i} 2852 & 1.444950 & 0.54 & 10.453 & 0.063 & 5.719 & 1.019 \\
Mg \textsc{ii} 2796, 2803 & 1.444950 & 0.50 & 11.837 & 0.137 & 4.155 & 0.638 \\
Mg \textsc{ii} 2796, 2803 & 1.446247 & 0.16 & 11.146 & 0.028 & 3.368 & 0.291 \\
Mg \textsc{ii} 2796, 2803 & 1.62936 & 2.49 & 11.432 & 0.122 & 11.167 & 3.630 \\
Mg \textsc{ii} 2796, 2803 & 1.629514 & 0.78 & 11.368 & 0.213 & 4.532 & 1.289 \\
Mg \textsc{ii} 2796, 2803 & 1.62983 & 1.26 & 12.404 & 0.045 & 18.429 & 1.008 \\
Fe \textsc{ii} 2344, 2382, 2600, 2586 & 1.629843 & 0.66 & 11.585 & 0.065 & 6.239 & 1.171 \\
Fe \textsc{ii} 2344, 2382, 2600, 2586 & 1.629902 & 0.26 & 11.999 & 0.262 & 4.115 & 0.426 \\
Mg \textsc{ii} 2796, 2803 & 1.629902 & 0.26 & 12.482 & 0.031 & 6.515 & 0.235 \\
Fe \textsc{ii} 2344, 2382, 2600, 2586 & 1.629969 & 0.17 & 11.957 & 0.038 & 5.428 & 0.373 \\
Mg \textsc{ii} 2796, 2803 & 1.629969 & 0.17 & 12.635 & 0.032 & 5.785 & 0.237 \\
Fe \textsc{ii} 2344, 2382, 2600 & 1.836536 & 0.19 & 11.287 & 0.050 & 1.942 & 0.414 \\
Fe \textsc{ii} 2344, 2382, 2600 & 1.8367040 & 0.08 & 11.833 & 0.015 & 2.129 & 0.138 \\
Fe \textsc{ii} 2344, 2382, 2600 & 1.836996 & 0.36 & 11.129 & 0.049 & 4.073 & 0.617 \\
Fe \textsc{ii} 2344, 2374, 2382 & 2.11649 & 1.03 & 11.289 & 0.608 & 2.271 & 2.386 \\
Fe \textsc{ii} 2344, 2374, 2382 & 2.116592 & 0.21 & 11.922 & 0.099 & 2.175 & 0.480 \\
Fe \textsc{ii} 2344, 2374, 2382 & 2.116676 & 0.53 & 12.754 & 0.029 & 10.737 & 0.735 \\
Fe \textsc{ii} 2344, 2374, 2382 & 2.116946 & 0.39 & 12.565 & 0.023 & 10.223 & 0.635 \\
Fe \textsc{ii} 2344, 2374, 2382 & 2.117289 & 0.15 & 12.367 & 0.051 & 1.717 & 0.254 \\
Fe \textsc{ii} 2344, 2374, 2382 & 2.117300 & 0.45 & 13.217 & 0.019 & 12.028 & 0.476 \\
Fe \textsc{ii} 2344, 2374, 2382 & 2.1174760 & 0.07 & 13.518 & 0.008 & 6.656 & 0.072 \\
Al \textsc{iii} 1854, 1862 & 2.300497 & 0.46 & 10.935 & 0.096 & 2.929 & 1.058 \\
Al \textsc{iii} 1854, 1862 & 2.300666 & 0.18 & 11.675 & 0.027 & 3.472 & 0.321 \\
Al \textsc{iii} 1854, 1862 & 2.301041 & 0.89 & 11.415 & 0.056 & 7.994 & 1.314 \\
Al \textsc{iii} 1854, 1862 & 2.302742 & 0.22 & 11.241 & 0.026 & 3.850 & 0.302 \\
Al \textsc{iii} 1854, 1862 & 2.303459 & 0.26 & 11.344 & 0.040 & 4.217 & 0.580 \\
Al \textsc{iii} 1854, 1862 & 2.304136 & 0.81 & 11.452 & 0.041 & 11.223 & 1.417 \\
C \textsc{iv} 1548, 1550 & 2.911425 & 0.51 & 11.659 & 0.192 & 4.204 & 1.100 \\
C \textsc{iv} 1548, 1550 & 2.911613 & 0.62 & 12.837 & 0.028 & 17.011 & 1.209 \\
C \textsc{iv} 1548, 1550 & 2.912090 & 0.31 & 12.255 & 0.014 & 8.846 & 0.349 \\
C \textsc{iv} 1548, 1550 & 2.912281 & 0.12 & 13.046 & 0.003 & 15.675 & 0.126 \\
C \textsc{iv} 1548, 1550 & 2.912303 & 0.68 & 11.867 & 0.033 & 8.160 & 0.740 \\
C \textsc{iv} 1548, 1550 & 2.912730 & 0.65 & 12.643 & 0.044 & 11.268 & 0.800 \\
C \textsc{iv} 1548, 1550 & 2.91318 & 2.43 & 11.895 & 0.401 & 9.656 & 3.912 \\
C \textsc{iv} 1548, 1550 & 2.91344 & 2.13 & 13.128 & 0.043 & 29.411 & 1.403 \\
C \textsc{iv} 1548, 1550 & 2.91354 & 1.15 & 12.653 & 0.114 & 9.038 & 1.399 \\
C \textsc{iv} 1548, 1550 & 2.913694 & 0.41 & 12.765 & 0.052 & 6.464 & 0.261 \\
C \textsc{iv} 1548, 1550 & 3.01060 & 5.27 & 12.941 & 0.114 & 20.110 & 3.115 \\
C \textsc{iv} 1548, 1550 & 3.01077 & 2.45 & 12.279 & 0.440 & 8.986 & 2.746 \\
C \textsc{iv} 1548, 1550 & 3.01097 & 2.73 & 12.475 & 0.136 & 11.209 & 2.337 \\
C \textsc{iv} 1548, 1550 & 3.12690 & 4.90 & 13.235 & 0.085 & 24.354 & 2.385 \\
C \textsc{iv} 1548, 1550 & 3.127032 & 0.77 & 12.571 & 0.135 & 9.130 & 0.926 \\
C \textsc{iv} 1548, 1550 & 3.12727 & 1.75 & 12.745 & 0.178 & 13.388 & 1.442 \\
C \textsc{iv} 1548, 1550 & 3.128909 & 0.40 & 12.394 & 0.022 & 9.451 & 0.630 \\
Si \textsc{ii} 1190, 1193, 1260, 1304, 1526 & 3.166417 & 0.27 & 12.409 & 0.021 & 5.363 & 0.356 \\
Al \textsc{ii} 1670 & 3.16647 & 1.12 & 11.755 & 0.066 & 12.811 & 1.746 \\
C \textsc{iv} 1548, 1550 & 3.166501 & 0.50 & 12.336 & 0.039 & 8.654 & 0.640 \\
Si \textsc{ii} 1526 & 3.166957 & 0.90 & 12.385 & 0.027 & 12.800 & 0.948 \\
Al \textsc{ii} 1670 & 3.166965 & 0.70 & 11.644 & 0.028 & 11.659 & 0.906 \\
C \textsc{iv} 1548, 1550 & 3.167111 & 0.57 & 13.350 & 0.006 & 37.359 & 0.718 \\
C \textsc{iv} 1548, 1550 & 3.16986 & 1.45 & 11.573 & 0.210 & 4.864 & 1.901 \\
C \textsc{iv} 1548, 1550 & 3.17016 & 1.00 & 12.503 & 0.056 & 9.966 & 1.039 \\
C \textsc{iv} 1548, 1550 & 3.17046 & 1.50 & 11.869 & 0.283 & 3.701 & 1.193 \\
C \textsc{iv} 1548, 1550 & 3.17056 & 1.73 & 12.181 & 0.186 & 5.507 & 1.534 \\
C \textsc{iv} 1548, 1550 & 3.170889 & 0.80 & 13.074 & 0.011 & 27.625 & 0.767 \\
C \textsc{iv} 1548, 1550 & 3.17093 & 1.43 & 12.202 & 0.232 & 6.461 & 1.482 \\
C \textsc{iv} 1548, 1550 & 3.171157 & 0.23 & 13.341 & 0.057 & 10.086 & 0.506 \\
C \textsc{iv} 1548, 1550 & 3.17184 & 1.57 & 12.710 & 0.026 & 26.687 & 2.057 \\
C \textsc{iv} 1548, 1550 & 3.17214 & 4.76 & 11.781 & 0.187 & 9.757 & 4.776 \\
C \textsc{iv} 1548, 1550 & 3.172948 & 0.54 & 12.485 & 0.030 & 8.993 & 0.599 \\
C \textsc{iv} 1548, 1550 & 3.173405 & 0.43 & 13.086 & 0.009 & 20.806 & 0.570 \\
C \textsc{iv} 1548, 1550 & 3.17397 & 1.02 & 12.143 & 0.059 & 7.459 & 1.141 \\
C \textsc{iv} 1548, 1550 & 3.17419 & 2.13 & 11.708 & 0.148 & 6.211 & 2.157 \\
C \textsc{iv} 1548, 1550 & 3.174690 & 0.80 & 12.861 & 0.034 & 10.919 & 0.649 \\
C \textsc{iv} 1548, 1550 & 3.174912 & 0.34 & 12.916 & 0.033 & 7.828 & 0.338 \\
C \textsc{iv} 1548, 1550 & 3.17536 & 2.35 & 13.294 & 0.068 & 22.115 & 2.337 \\
C \textsc{iv} 1548, 1550 & 3.175694 & 0.68 & 12.957 & 0.107 & 11.871 & 1.333 \\
C \textsc{iv} 1548, 1550 & 3.17596 & 1.49 & 12.537 & 0.155 & 10.011 & 1.851 \\
C \textsc{iv} 1548, 1550 & 3.28575 & 1.64 & 12.219 & 0.060 & 13.116 & 2.019 \\
C \textsc{iv} 1548, 1550 & 3.28617 & 1.10 & 12.582 & 0.059 & 13.463 & 1.357 \\
C \textsc{iv} 1548, 1550 & 3.286672 & 0.74 & 13.027 & 0.032 & 21.250 & 1.770 \\
C \textsc{iv} 1548, 1550 & 3.287079 & 0.51 & 12.896 & 0.040 & 9.966 & 0.577 \\
C \textsc{iv} 1548, 1550 & 3.28733 & 1.18 & 12.543 & 0.058 & 9.895 & 1.105 \\
C \textsc{iv} 1548, 1550 & 3.293307 & 0.28 & 12.504 & 0.013 & 8.029 & 0.278 \\
Si \textsc{iv} 1393, 1402 & 3.334303 & 0.32 & 11.418 & 0.119 & 4.044 & 0.633 \\
C \textsc{iv} 1548, 1550 & 3.334325 & 0.50 & 12.293 & 0.213 & 8.369 & 1.540 \\
Si \textsc{iv} 1393, 1402 & 3.33456 & 1.47 & 12.551 & 0.051 & 18.584 & 2.044 \\
C \textsc{iv} 1548, 1550 & 3.33461 & 1.49 & 12.978 & 0.089 & 21.013 & 4.118 \\
Si \textsc{iv} 1393, 1402 & 3.334662 & 0.50 & 10.962 & 0.159 & 2.658 & 0.778 \\
Si \textsc{iv} 1393, 1402 & 3.334951 & 0.54 & 11.738 & 0.017 & 11.749 & 0.525 \\
Si \textsc{iv} 1393, 1402 & 3.335219 & 0.44 & 12.281 & 0.056 & 8.025 & 0.496 \\
C \textsc{iv} 1548, 1550 & 3.335228 & 0.57 & 12.617 & 0.013 & 16.343 & 0.557 \\
Al \textsc{ii} 1670 & 3.335367 & 0.12 & 11.603 & 0.026 & 4.390 & 0.188 \\
Si \textsc{ii} 1526 & 3.335367 & 0.12 & 12.805 & 0.021 & 4.057 & 0.141 \\
Si \textsc{iv} 1393, 1402 & 3.335380 & 0.50 & 11.287 & 0.166 & 3.373 & 0.714 \\
C \textsc{iv} 1548, 1550 & 3.33548 & 4.42 & 12.443 & 0.572 & 12.199 & 3.604 \\
Si \textsc{iv} 1393, 1402 & 3.335490 & 0.96 & 12.645 & 0.027 & 14.589 & 0.767 \\
Al \textsc{ii} 1670 & 3.335532 & 0.52 & 11.713 & 0.024 & 9.875 & 0.584 \\
Si \textsc{ii} 1526 & 3.335532 & 0.52 & 12.858 & 0.021 & 9.309 & 0.488 \\
C \textsc{iv} 1548, 1550 & 3.335914 & 0.70 & 12.750 & 0.023 & 16.190 & 0.990 \\
Si \textsc{iv} 1393, 1402 & 3.335956 & 0.73 & 11.851 & 0.045 & 10.186 & 1.149 \\
C \textsc{iv} 1548, 1550 & 3.38587 & 1.37 & 11.927 & 0.314 & 5.616 & 4.883 \\
C \textsc{iv} 1548, 1550 & 3.486931 & 0.66 & 12.370 & 0.136 & 7.867 & 1.419 \\
Si \textsc{iv} 1393, 1402 & 3.487848 & 0.34 & 11.988 & 0.041 & 5.634 & 0.485 \\
C \textsc{iv} 1548, 1550 & 3.487848 & 0.68 & 12.278 & 0.106 & 5.901 & 0.868 \\
C \textsc{iv} 1548, 1550 & 3.48814 & 6.41 & 13.283 & 0.083 & 23.026 & 2.761 \\
Si \textsc{iv} 1393, 1402 & 3.48815 & 1.56 & 12.607 & 0.055 & 12.240 & 1.181 \\
Si \textsc{iv} 1393, 1402 & 3.488320 & 0.22 & 11.953 & 0.089 & 2.990 & 0.384 \\
C \textsc{iv} 1548, 1550 & 3.488322 & 0.87 & 12.320 & 0.676 & 4.645 & 2.099 \\
Si \textsc{iv} 1393, 1402 & 3.488415 & 0.80 & 12.817 & 0.039 & 9.322 & 0.388 \\
C \textsc{iv} 1548, 1550 & 3.48842 & 4.79 & 12.908 & 0.345 & 8.884 & 3.234 \\
C \textsc{iv} 1548, 1550 & 3.48984 & 1.39 & 12.611 & 0.051 & 12.046 & 1.563 \\
Si \textsc{iv} 1393, 1402 & 3.489858 & 0.65 & 11.669 & 0.033 & 7.802 & 0.804 \\
C \textsc{iv} 1548, 1550 & 3.49049 & 2.43 & 12.413 & 0.129 & 14.553 & 5.219 \\
C \textsc{iv} 1548, 1550 & 3.58757 & 2.47 & 12.348 & 0.045 & 20.347 & 2.494 \\
C \textsc{iv} 1548, 1550 & 3.588900 & 0.48 & 12.321 & 0.153 & 7.428 & 1.122 \\
C \textsc{iv} 1548, 1550 & 3.58904 & 1.49 & 12.897 & 0.015 & 38.239 & 1.613 \\
C \textsc{iv} 1548, 1550 & 3.58973 & 1.93 & 12.001 & 0.052 & 13.181 & 1.942 \\
C \textsc{iv} 1548, 1550 & 3.58978 & 4.23 & 12.027 & 0.092 & 16.131 & 3.930 \\
C \textsc{iv} 1548, 1550 & 3.59860 & 1.42 & 12.157 & 0.065 & 13.409 & 2.583 \\
C \textsc{iv} 1548, 1550 & 3.60138 & 2.31 & 12.744 & 0.108 & 11.691 & 1.222 \\
C \textsc{iv} 1548, 1550 & 3.60177 & 3.61 & 13.323 & 0.180 & 15.903 & 2.943 \\
Si \textsc{iv} 1393, 1402 & 3.601972 & 0.28 & 11.950 & 0.022 & 5.156 & 0.341 \\
C \textsc{iv} 1548, 1550 & 3.601976 & 0.19 & 12.725 & 0.064 & 5.488 & 0.354 \\
C \textsc{iv} 1548, 1550 & 3.60213 & 1.08 & 12.632 & 0.021 & 19.093 & 1.114 \\
C \textsc{iv} 1548, 1550 & 3.603594 & 0.40 & 12.123 & 0.039 & 4.698 & 0.607 \\
C \textsc{ii} 1334 & 3.63199 & 1.59 & 12.430 & 0.049 & 11.481 & 1.378 \\
Fe \textsc{ii} 1608 & 3.63211 & 1.23 & 12.715 & 0.037 & 14.608 & 1.681 \\
C \textsc{ii} 1334 & 3.632238 & 0.13 & 12.837 & 0.006 & 6.598 & 0.115 \\
O \textsc{i} 1302 & 3.63224 & 1.12 & 12.802 & 0.117 & 7.089 & 1.428 \\
Si \textsc{iv} 1393, 1402 & 3.63227 & 2.43 & 12.685 & 0.027 & 32.986 & 2.545 \\
C \textsc{iv} 1548, 1550 & 3.63228 & 1.67 & 12.898 & 0.039 & 15.635 & 1.708 \\
Al \textsc{ii} 1670 & 3.63230 & 1.35 & 11.591 & 0.147 & 11.493 & 1.855 \\
Si \textsc{iv} 1393, 1402 & 3.632306 & 0.31 & 12.424 & 0.032 & 8.032 & 0.354 \\
Si \textsc{ii} 1304, 1526 & 3.63230 & 1.52 & 12.466 & 0.045 & 11.605 & 1.119 \\
C \textsc{ii} 1334 & 3.632341 & 0.25 & 12.992 & 0.006 & 13.940 & 0.224 \\
C \textsc{ii} 1334 & 3.632393 & 0.61 & 12.373 & 0.261 & 4.389 & 0.843 \\
Si \textsc{ii} 1304, 1526 & 3.632394 & 0.66 & 11.236 & 0.256 & 1.320 & 1.081 \\
O \textsc{i} 1302 & 3.632396 & 0.75 & 12.659 & 0.171 & 4.616 & 0.796 \\
O \textsc{i} 1302 & 3.63261 & 3.06 & 13.376 & 0.075 & 18.343 & 3.556 \\
Si \textsc{ii} 1304, 1526 & 3.632720 & 0.30 & 11.990 & 0.048 & 2.815 & 0.400 \\
O \textsc{i} 1302 & 3.632723 & 0.10 & 13.121 & 0.023 & 3.803 & 0.127 \\
C \textsc{ii} 1334 & 3.632727 & 0.11 & 12.823 & 0.016 & 4.454 & 0.115 \\
Al \textsc{ii} 1670 & 3.632753 & 0.65 & 10.624 & 0.193 & 2.666 & 0.999 \\
C \textsc{ii} 1334 & 3.63296 & 1.04 & 12.304 & 0.096 & 5.001 & 1.009 \\
O \textsc{i} 1302 & 3.632977 & 0.69 & 12.488 & 0.181 & 5.160 & 1.023 \\
C \textsc{ii} 1334 & 3.63311 & 1.04 & 12.284 & 0.186 & 4.823 & 1.492 \\
Al \textsc{ii} 1670 & 3.63315 & 6.85 & 11.122 & 0.100 & 16.002 & 3.569 \\
Si \textsc{iv} 1393, 1402 & 3.633185 & 0.61 & 12.506 & 0.028 & 8.949 & 0.556 \\
C \textsc{iv} 1548, 1550 & 3.63318 & 1.78 & 13.084 & 0.060 & 12.195 & 1.070 \\
C \textsc{ii} 1334 & 3.63326 & 1.10 & 12.477 & 0.095 & 5.882 & 1.091 \\
O \textsc{i} 1302 & 3.633274 & 0.30 & 12.506 & 0.076 & 3.068 & 0.351 \\
O \textsc{i} 1302 & 3.633357 & 0.80 & 13.066 & 0.013 & 19.848 & 0.680 \\
C \textsc{iv} 1548, 1550 & 3.633494 & 0.12 & 13.416 & 0.044 & 6.122 & 0.249 \\
Si \textsc{iv} 1393, 1402 & 3.6335000 & 0.08 & 13.032 & 0.024 & 5.079 & 0.139 \\
Si \textsc{ii} 1304, 1526 & 3.633510 & 0.51 & 11.478 & 0.124 & 1.025 & 0.951 \\
C \textsc{ii} 1334 & 3.633522 & 0.45 & 13.077 & 0.029 & 7.407 & 0.448 \\
O \textsc{i} 1302 & 3.63352 & 1.06 & 13.157 & 0.267 & 6.425 & 1.284 \\
Al \textsc{ii} 1670 & 3.633538 & 0.70 & 11.263 & 0.151 & 5.338 & 0.858 \\
C \textsc{iv} 1548, 1550 & 3.63360 & 2.02 & 13.519 & 0.055 & 14.129 & 0.752 \\
Si \textsc{iv} 1393, 1402 & 3.633609 & 0.87 & 13.073 & 0.027 & 12.083 & 0.382 \\
O \textsc{i} 1302 & 3.633681 & 0.72 & 12.837 & 0.519 & 3.126 & 0.960 \\
Si \textsc{ii} 1304, 1526 & 3.63368 & 2.09 & 12.384 & 0.211 & 8.106 & 3.056 \\
Al \textsc{ii} 1670 & 3.633699 & 0.48 & 11.208 & 0.127 & 4.574 & 0.611 \\
C \textsc{ii} 1334 & 3.633699 & 0.27 & 13.025 & 0.041 & 5.631 & 0.347 \\
O \textsc{i} 1302 & 3.633766 & 0.12 & 13.069 & 0.007 & 5.436 & 0.111 \\
Al \textsc{ii} 1670 & 3.633880 & 0.75 & 11.268 & 0.349 & 5.869 & 2.170 \\
C \textsc{ii} 1334 & 3.633884 & 0.13 & 13.415 & 0.009 & 6.534 & 0.130 \\
Fe \textsc{ii} 1608 & 3.633888 & 0.62 & 12.275 & 0.076 & 3.477 & 0.955 \\
O \textsc{i} 1302 & 3.63389 & 1.21 & 13.602 & 0.106 & 5.683 & 0.421 \\
Si \textsc{ii} 1304, 1526 & 3.633891 & 0.42 & 12.630 & 0.037 & 5.495 & 0.305 \\
C \textsc{ii} 1334 & 3.634066 & 0.84 & 11.629 & 0.136 & 2.893 & 1.038 \\
C \textsc{iv} 1548, 1550 & 3.66797 & 1.06 & 12.558 & 0.079 & 10.854 & 2.198 \\
C \textsc{iv} 1548, 1550 & 3.668310 & 0.83 & 12.507 & 0.043 & 8.492 & 0.724 \\
Si \textsc{iv} 1393, 1402 & 3.707394 & 0.41 & 11.773 & 0.047 & 4.329 & 0.362 \\
Si \textsc{iv} 1393, 1402 & 3.707592 & 0.45 & 12.134 & 0.110 & 8.217 & 0.910 \\
C \textsc{iv} 1548, 1550 & 3.707634 & 0.73 & 12.769 & 0.015 & 17.691 & 0.749 \\
Si \textsc{iv} 1393, 1402 & 3.70787 & 4.65 & 12.311 & 0.099 & 20.498 & 4.312 \\
C \textsc{iv} 1548, 1550 & 3.70800 & 1.54 & 13.297 & 0.018 & 28.860 & 0.955 \\
C \textsc{iv} 1548, 1550 & 3.70869 & 1.59 & 12.557 & 0.094 & 13.005 & 1.509 \\
C \textsc{iv} 1548, 1550 & 3.78479 & 1.06 & 12.856 & 0.023 & 15.534 & 0.612 \\
Si \textsc{iv} 1393, 1402 & 3.784828 & 0.33 & 12.043 & 0.017 & 7.777 & 0.399 \\
C \textsc{iv} 1548, 1550 & 3.784896 & 0.20 & 12.570 & 0.034 & 4.906 & 0.254 \\
C \textsc{iv} 1548, 1550 & 3.79516 & 2.10 & 12.255 & 0.128 & 8.989 & 2.560 \\
C \textsc{iv} 1548, 1550 & 3.79568 & 3.99 & 12.990 & 0.096 & 19.425 & 3.161 \\
C \textsc{iv} 1548, 1550 & 3.79612 & 4.27 & 12.173 & 0.100 & 13.856 & 3.487 \\
C \textsc{iv} 1548, 1550 & 3.837585 & 0.80 & 11.889 & 0.254 & 4.680 & 1.432 \\
Si \textsc{iv} 1393, 1402 & 3.837603 & 0.74 & 11.516 & 0.042 & 5.939 & 0.741 \\
C \textsc{iv} 1548, 1550 & 3.83779 & 2.14 & 12.744 & 0.072 & 16.676 & 2.851 \\
C \textsc{iv} 1548, 1550 & 3.83816 & 1.35 & 11.995 & 0.152 & 7.026 & 1.576 \\
C \textsc{iv} 1548, 1550 & 3.83867 & 6.33 & 12.941 & 0.171 & 15.501 & 2.524 \\
Si \textsc{iv} 1393, 1402 & 3.838675 & 0.68 & 12.010 & 0.030 & 10.330 & 0.962 \\
C \textsc{iv} 1548, 1550 & 3.839101 & 0.32 & 12.754 & 0.010 & 11.559 & 0.316 \\
C \textsc{iv} 1548, 1550 & 3.84380 & 1.15 & 11.899 & 0.056 & 7.427 & 1.233 \\
C \textsc{iv} 1548, 1550 & 3.845443 & 0.43 & 12.598 & 0.018 & 9.369 & 0.453 \\
C \textsc{iv} 1548, 1550 & 3.84625 & 1.04 & 12.648 & 0.021 & 16.865 & 0.979 \\
C \textsc{iv} 1548, 1550 & 3.84683 & 3.12 & 11.904 & 0.099 & 10.730 & 2.798 \\
C \textsc{iv} 1548, 1550 & 3.862214 & 0.56 & 12.339 & 0.028 & 9.058 & 0.798 \\
C \textsc{iv} 1548, 1550 & 3.897837 & 0.29 & 12.856 & 0.007 & 14.606 & 0.300 \\
C \textsc{iv} 1548, 1550 & 3.90051 & 2.39 & 12.463 & 0.109 & 21.360 & 6.125 \\
C \textsc{iv} 1548, 1550 & 3.90220 & 2.63 & 12.319 & 0.091 & 9.651 & 1.526 \\
C \textsc{iv} 1548, 1550 & 3.902256 & 0.77 & 11.204 & 0.372 & 1.000 & 1.738 \\
C \textsc{iv} 1548, 1550 & 3.903427 & 0.60 & 12.320 & 0.033 & 7.570 & 0.802 \\
C \textsc{iv} 1548, 1550 & 3.904377 & 0.57 & 12.367 & 0.026 & 8.343 & 0.669 \\
C \textsc{iv} 1548, 1550 & 3.904716 & 0.15 & 12.953 & 0.006 & 7.675 & 0.138 \\
\end{longtable}

\end{document}